\documentclass[smallextended]{svjour3}
\usepackage[T1]{fontenc}
\usepackage{lmodern}
\usepackage{amsmath}
\usepackage{mathptmx}
\usepackage{graphicx}
\usepackage{color}
\usepackage{array}
\usepackage[a4paper]{geometry}

\def\diffd{\mathrm{d}}
\DeclareMathOperator{\proba}{Proba}
\def\AND{\text{ and }}

\journalname{}
\title{A branching random walk seen from the tip}
\author{\'Eric Brunet \and Bernard Derrida}
\institute{Laboratoire de Physique Statistique, \'Ecole Normale
Sup\'erieure, UPMC, Universit\'e Paris Diderot, CNRS, 24 rue Lhomond,
75005 Paris,
France.\\\email{Eric.Brunet@lps.ens.fr}\\\email{Bernad.Derrida@lps.ens.fr}}
\date{\today}

\begin{document}

\maketitle

\begin{abstract}
We show that all the time-dependent statistical properties of the rightmost
points of a branching Brownian motion can be extracted from the traveling
wave solutions of the Fisher-KPP equation. We show that the distribution of
all the distances between the rightmost points has a long time limit which
can be understood as the delay of the Fisher-KPP traveling waves when the
initial condition is modified. The limiting
measure exhibits the surprising property of superposability: the
statistical properties of the distances between the rightmost points of the
union of two realizations of the branching Brownian motion shifted by
arbitrary amounts are the same as those of a single realization. We discuss
the extension of our results to more general branching random walks.

\PACS{02.50.-r \and 05.40.-a \and 89.75.Hc}
\end{abstract}

\section{Introduction}

\begin{figure}[h]
\centering
\includegraphics[width=.35\textwidth]{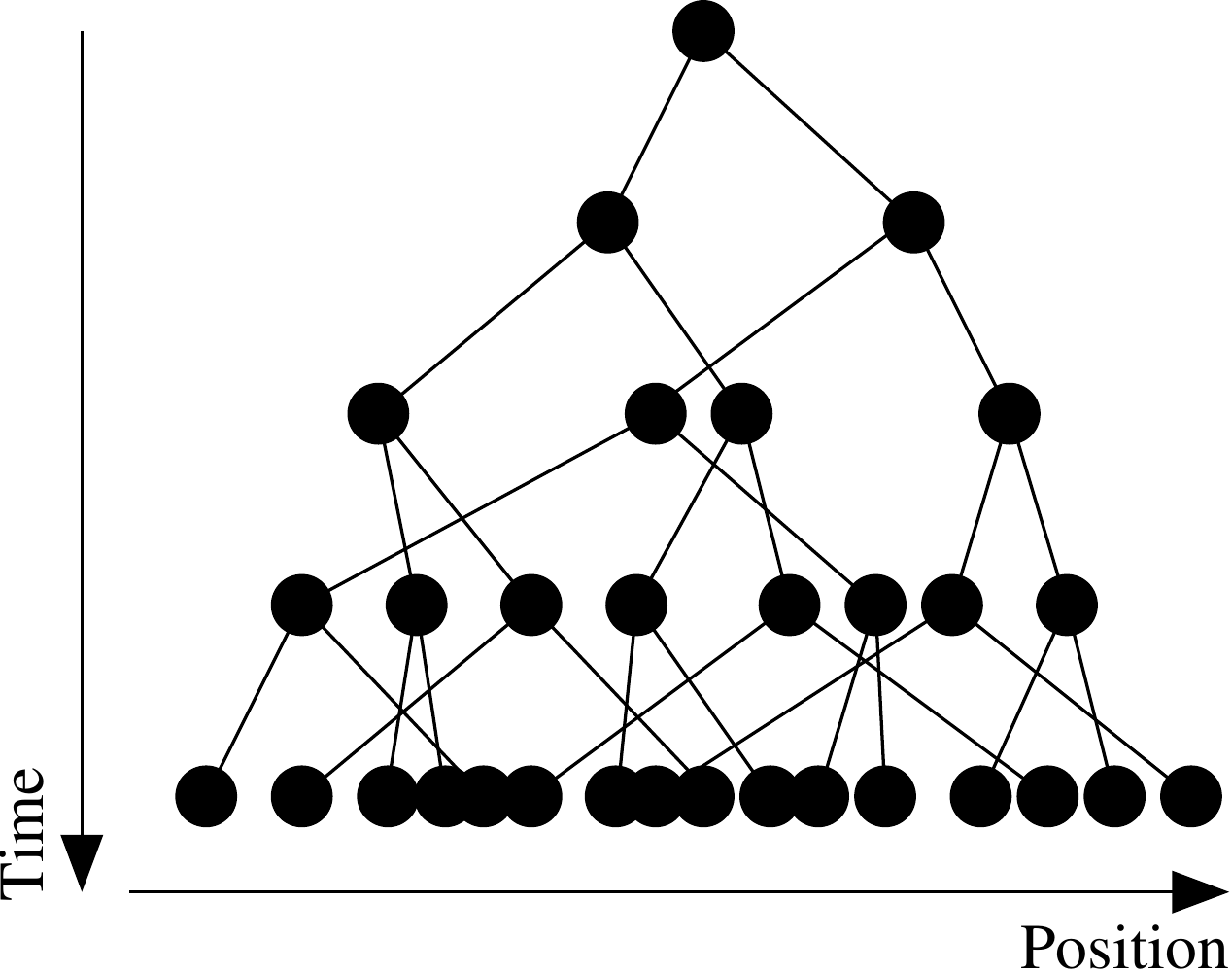}
\qquad\qquad
\includegraphics[width=.35\textwidth]{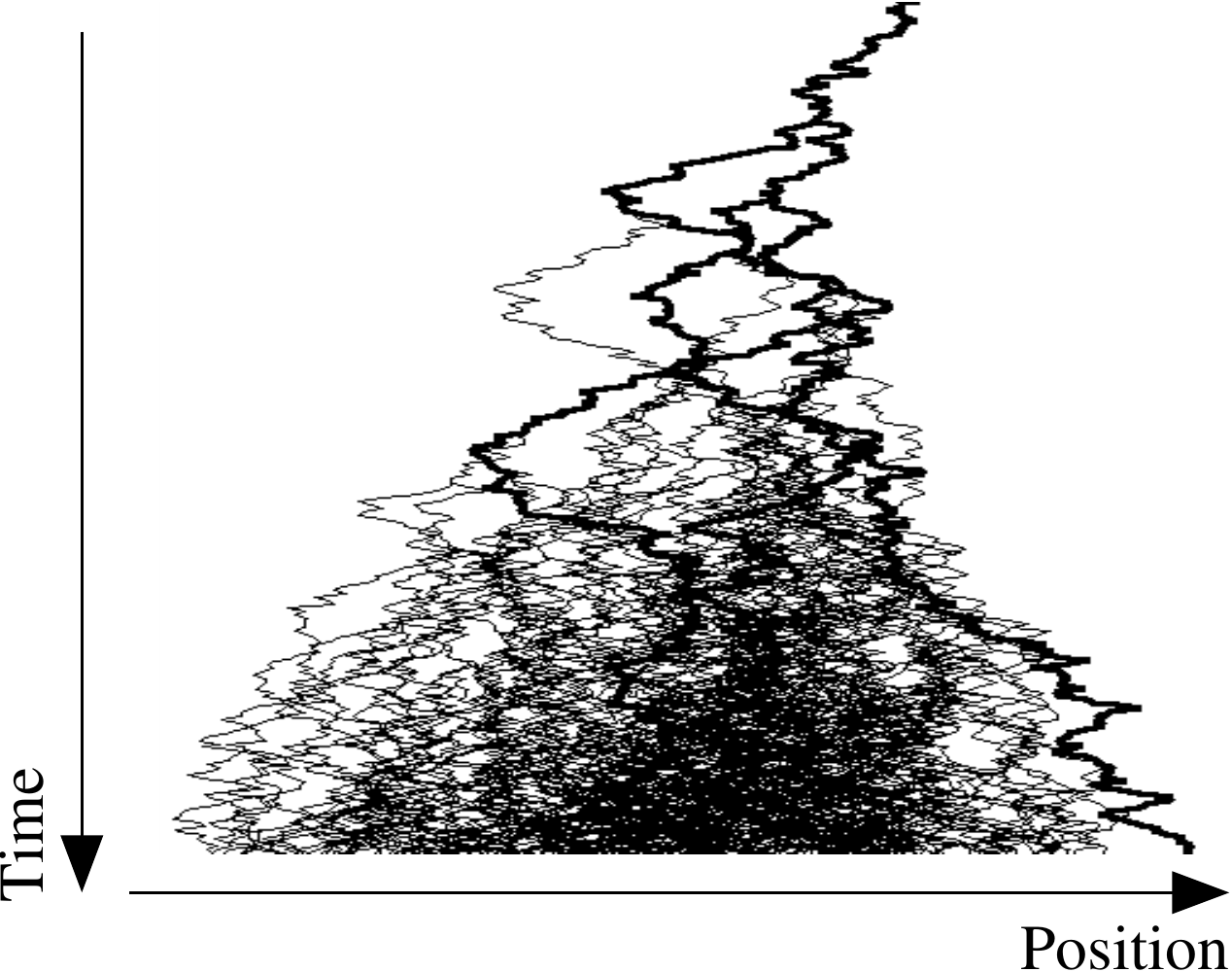}
\caption{Two examples of branching random walks. Left: a branching random
walk with discrete time where each point splits into two points at each
time step. Right: a continuous version called branching Brownian motion
where points diffuse as in a Brownian motion and branch with a constant rate.}
\label{fig:brw}
\end{figure}

A branching random walk is a collection of points which, starting from
a single point, diffuse and branch independently of the time, of their
positions or of the other points, as in figure~\ref{fig:brw}.

Branching random walks appear in many contexts ranging from Mathematics
\cite{McKean.75,Bramson.83,LalleySellke.87} to Biology
\cite{Fisher.37,Kessler.97,GoldingKozlovsky.98}. They can for example be
used to describe how a growing population invades a new environment. In the
one dimensional case, see figure~\ref{fig:brw}, there is, at a given time
$t$, a rightmost individual at position $X_1(t)$, a second rightmost at
$X_2(t)$ and so on. (Note that the rightmost $X_1(t')$ at a time $t'>t$ is
not necessarily a descendant of the rightmost $X_1(t)$ at time $t$.) The
expected position $m_t=\langle X_1(t)\rangle$ of the rightmost individual
as well as the probability distribution of its position $X_1(t)$ around
$m_t$ are well understood \cite{Bramson.83,McKean.75}; the goal of the
present paper is to describe the statistical properties of the positions
of all the rightmost points in the system, in particular the distribution
of the distances between the two rightmost points, the average density of
points at some fixed distance from the rightmost $X_1(t)$, etc.

One motivation for studying these distances is that the problem belongs to
the broader context of extreme value statistics
\cite{TracyWidom.94,BouchaudMezard.97,DeanMajumdar.01,MajumdarKrapivsky.02,SabhapanditMajumdar.07,Aizenman.05,Burkhardt.07,Gyorgyi.08}:
Trying to understand the statistical properties of the rightmost points in
a random set of points on the line is a problem common to the studies of
the largest eigenvalues of random matrices \cite{TracyWidom.94}, of the
extrema of random signals
\cite{Derrida.81,Derrida.85,Ruelle.87,BolthausenSznitman.98,IgloiMonthus.05,BovierKurkova.06},
or of the low lying states of some disordered systems such as spin glasses
\cite{MPSTV.84,Aizenman.05,BovierKurkova.07,Aizenman.07,Arguin.07}. In fact, the points
generated after some time $t$ by a branching random walk can be viewed as
the energies of the configurations of a directed polymer in a random medium
\cite{DerridaSpohn.88,MonthusGarel.07}, and the distances between the rightmost points as
the gaps between the low lying energy states.

The most studied example of branching random walk is the branching Brownian
motion: one starts with a single point at the origin which performs
a Brownian motion and branches at a given fixed rate (right part of
figure~\ref{fig:brw}). Whenever a branching event occurs, the point is
replaced by two new points which evolve themselves as two independent
branching Brownian motions. While the number of points generated after some
time $t$ grows exponentially with time, the expected position $m_t$ of the
rightmost point increases only linearly with time
\cite{McKean.75,Bramson.83}. In one dimension, Mc Kean \cite{McKean.75} and
Bramson \cite{Bramson.83} have shown that the probability distribution of
the rightmost point is given by the traveling wave solution of the Fisher-KPP
equation, with a step initial condition. Here we will see that all the
statistical properties of the rightmost points can be understood in terms
of solutions to the Fisher-KPP equation with appropriate initial conditions
\cite{BrunetDerrida.09}. We will also show that the distribution of the
distances between these rightmost points has a long time limit which
exhibits the striking property of superposability: the distances between
the rightmost points of the union of two realizations of the branching
Brownian motion have the same statistics as those of a single realization.

This paper is organized as follows: in section~\ref{statistics} we
introduce some generating functions useful to study random sets of points
on the line and show how one can use them to obtain all the properties of
these random sets. In section~\ref{sec:BBMFKPP} we show that, for the
branching Brownian motion, all these generating functions are solutions of
the Fisher-KPP equation. We also show that the distribution of all the
rightmost points as seen from $m_t$ or, alternatively, as seen from
$X_1(t)$, has a long time limit which can be computed as the delay of
Fisher-KPP traveling waves. This distribution has the property of
superposability. In section~\ref{quantitative}, we present results, mostly
numerical, on some specific aspects of the limiting distribution of points
in the branching Brownian motion, namely the distribution of the distance
between the two rightmost points and the average density seen from the
rightmost point. In section~\ref{sec:disc} we explain how the results on
the branching Brownian motion can be extended to more general branching
random walks. Finally, we study in section~\ref{statmeas} the distribution
of all the rightmost points in a specific frame which depends on the
realization and which was introduced by Lalley and Sellke
\cite{LalleySellke.87}.

\section{Statistics of point measures on the line} \label{statistics}

In this section, we introduce some useful quantities (generating functions)
to characterize random sets of points on the line such that the number
$n(x)$ defined as
\begin{equation}
n(x)=\text{(the number of points on  the right of position $x$)}
\end{equation}
is finite and vanishes for $x$ large enough.

\subsection{The generating functions} \label{generating-functions}

The first generating function one can define is 
\begin{equation}
\psi_\lambda(x)=\big\langle \lambda^{n(x)} \big\rangle.
\label{defpsi}
\end{equation}
From the knowledge of this function, one can extract the 
probability distribution function $p_i(x)$ of the position $x$ of the $i$-th
rightmost point. Indeed, by definition \eqref{defpsi} of
$\psi_\lambda$,
\begin{equation}
\psi_\lambda(x)=\sum_{i\ge0} Q_i(x) \lambda^i,
\label{exppsi}
\end{equation}
where $Q_i(x)$ is the probability that there are exactly $i$ points on
the right of $x$.
One can notice that
$Q_0(x)+Q_1(x)+\cdots + Q_{i-1}(x)$ is the probability to have less than
$i$ points on the right of $x$. The generating function of these sums is,
from \eqref{exppsi},
\begin{equation}
\frac \lambda {1-\lambda}\psi_\lambda(x)=
Q_0(x)\lambda+\big[Q_0(x)+Q_1(x)\big]\lambda^2+\big[Q_0(x)+Q_1(x)+Q_2(x)\big]\lambda^3+\cdots.
\label{sumterms}
\end{equation}
But $Q_0(x)+Q_1(x)+\cdots + Q_{i-1}(x)$ is also the probability that
the $i$-th rightmost point, if it exists, is on the left of $x$.
Therefore,
\begin{equation}
\frac\lambda{1-\lambda}\partial_x\psi_\lambda(x)=\sum_{i\ge1}p_i(x)\lambda^i,
\label{pix}
\end{equation}
where $p_i(x)\,\diffd x$ is the probability that the $i$-th
rightmost point exists and is in the interval $[x,x+\diffd x]$.
(Note that $\int p_i(x)\,\diffd x\le 1$ is the probability that there are at
least $i$ points on the line.)

The knowledge of $\psi_\lambda(x)$ gives in particular the average distances
between the points: from (\ref{pix}), one can see that
\begin{align*}
\int \diffd x\, x \, \partial_x \psi_\lambda(x)
&= (1-\lambda) \big[ \langle X_1 \rangle + \lambda \langle X_2 \rangle
+ \lambda^2 \langle X_3 \rangle + \cdots\big ],\notag\\
&=\langle X_1 \rangle-\lambda[\langle X_1 \rangle-\langle X_2 \rangle]
-\lambda^2[\langle X_2 \rangle-\langle X_3 \rangle]-\cdots,
\end{align*}
where $\langle X_i \rangle=\int x p_i(x)\,\diffd x$ is the average position
of the $i$-th point (with the convention that $X_i=0$ if there are less
than $i$ points in the system). Therefore
\begin{equation}
\int \diffd x\, x \left[\partial_x \psi_0(x)-\partial_x \psi_\lambda(x)\right]
=\sum_{i\ge1} \langle d_{i,i+1}\rangle \lambda^i,
\label{psid}
\end{equation}
where $\langle d_{i,i+1}\rangle =\langle X_i\rangle-\langle X_{i+1}\rangle$
is the average distance between the $i$-th and the $(i+1)$-th point.

To obtain the correlations between the positions of pairs of points, one
can start, for $y < x$, from the generating function
\begin{equation}
\psi_{\lambda\mu}(x,y) = \big\langle \lambda^{n(x)}\mu^{n(y)} \big\rangle.
\label{defpsi2}
\end{equation}
The coefficient in front of $\lambda^i \mu^j$ in the expansion of
$\psi_{\lambda\mu}$ in powers of $\lambda$ and $\mu$ is the probability
that there are exactly $i$ points on the right of $x$ and $j$ points on the
right of $y$. As in \eqref{sumterms}, the coefficient of $\lambda^i \mu^j$
in the expansion of $\lambda/(1-\lambda) \times \mu/(1-\mu) \times
\psi_{\lambda\mu}(x,y)$ is the probability that there are less than $i$
points on the right of $x$ and less than $j$ points on the right of $y$,
which is also the probability that the $i$-th rightmost point (if it
exists) is on the left of $x$ and the $j$-th rightmost point (if it exists)
is on the left of $y$. Thus, for $y<x$,
\begin{equation}
\frac\lambda{1-\lambda}\frac\mu{1-\mu}\partial_x\partial_y\psi_{\lambda\mu}(x,y)
= \sum_{\substack{i\ge1\\j>i}}p_{ij}(x,y) \lambda^i\mu^j,
\label{twopoint}
\end{equation}
where $p_{ij}(x,y)\,\diffd x\,\diffd y$ is the probability that both the
$i$-th and $j$-th rightmost points exist and lie respectively in the
intervals $[x,x+\diffd x]$ and $[y,y+\diffd y]$.

One can generalize (\ref{defpsi},\ref{defpsi2}) by defining, for $x_0 > x_1
> \cdots > x_k$, the generating functions
\begin{equation}
\psi_{\lambda_0, \ldots, \lambda_k}(x_0, \ldots, x_k)
 = \left\langle \lambda_0^{n(x_0)} \cdots \lambda_k^{n(x_k)}\right\rangle 
\label{defpsi3}
\end{equation}
of the numbers $n(x_0), \ldots, n(x_k)$ of points on the right of positions
$x_0,\ldots, x_k$, and get as in (\ref{pix},\ref{twopoint}) all the higher
correlation functions. In that way, all the statistical properties of the
measure can be derived from the knowledge of the generating functions
(\ref{defpsi3}).

\subsection{The measure seen from the rightmost point}
In the following we will often try to characterize the random set of
points as seen from the rightmost point (i.e. in the frame where the
rightmost point is at the origin). To do so, let us define the generating
functions of the numbers $m(z)$ of points at the right of $z$ in the frame
of the rightmost point. (Note that if $X_1$ is the position of the rightmost,
then $m(z)=n(X_1+z)$ and one has $m(z) \ge 1$ for $z <0$ and $m(z)=0$ for
$z >0$.)
\begin{equation}
\chi_{\lambda_1, \ldots, \lambda_k}(z_1, \ldots, z_k) = \left\langle \lambda_1^{m(z_1)} \cdots \lambda_k^{m(z_k)}\right\rangle. 
\label{defchi3}
\end{equation}
(As in \eqref{defpsi3}, we assume $z_1>z_2>\cdots>z_k$.) These generating
functions, as in section \ref{generating-functions}, allow one to calculate
all the statistical properties of the measure in the frame of the rightmost
point (in particular the distribution of the relative distances between the
points). They can be determined from the knowledge of the generating
functions $\psi_{\lambda_0,\ldots, \lambda_k}(x_0, \ldots, x_k)$ defined in
(\ref{defpsi3}) by
\begin{equation}
\chi_{\lambda_1,\ldots, \lambda_k}(z_1, \ldots, z_k) =
\int \diffd x\, \partial_{x_0} \psi_{0,\lambda_1, \ldots,\lambda_k}
(x,x+z_1, \ldots, x+z_k).
\label{the-frame}
\end{equation}

In section \ref{quantitative} we will calculate the density of probability
$P_{12}(a)$ that the two rightmost points are separated by
a distance $a$ (and that there are at least two points on the line) and
the average density $\rho(a)$ at a distance $a$ from the rightmost point.
From
(\ref{defchi3}) one can see that
\begin{equation}
P_{12}(a)= - \partial_a \partial_{\mu} \chi_\mu(-a) \Big|_{\mu=0}
= - \partial_a \int \diffd x \, \partial_{\mu} \partial_{x_0}
 \psi_{0\mu}(x,x-a) \Big|_{\mu=0} .
\end{equation}
Then using that $ \partial_{x_0} \psi_{0\mu}(x,x-a) = ( \partial_{x}
+ \partial_{a} ) \psi_{0\mu}(x,x-a)$, one gets
\begin{equation}
P_{12}(a)= - 
\partial_a^2\int \diffd x\,\partial_{\mu} 
\psi_{0\mu}(x,x-a)\Big|_{\mu=0} .
\label{P12}
\end{equation}
By a similar calculation one can show that the average density $\rho(a)$
of points at distance $a$ from the rightmost point is 
\begin{equation}
\rho(a)= 
\partial_a^2\int \diffd x\,\partial_{\mu}
\psi_{0\mu}(x,x-a)\Big|_{\mu=1}.
\label{rho}
\end{equation}

\subsection{Examples}

We now describe a few examples of such measures.

\subsubsection{A Poisson process with an arbitrary density $r(x)$} \label{Poisson Process}

Our first example is a Poisson process on the line with a density $r(x)$.
We assume that $r(x)$ decays fast enough to the right so that a rightmost
point exists, and that $\int r(x)\, \diffd x= \infty$ so that there are
infinitely many points on the line.

By definition of a Poisson process, each infinitesimal interval
$[x,x+\diffd x]$ is occupied by a point with probability $r(x)\, \diffd x$
 and empty with probability $1-r(x)\,\diffd x$, and the occupation numbers of
disjoint intervals are uncorrelated. The probability $Q_i(x)$ that there
are exactly $i$ points on the right of $x$ is given by
\begin{equation}
Q_i(x) = \frac{R(x)^i e^{- R(x)}} {i !}
\qquad\text{where}\qquad
R(x)= \int_x^\infty r(z)\, \diffd z.
\end{equation}
From this, we obtain $\psi_\lambda(x)$ from (\ref{defpsi},\ref{exppsi}) and
$\psi_{\lambda\mu}(x,y)$ from (\ref{defpsi2}) in the Poisson process:
\begin{equation}
\psi_\lambda(x)= e^{-(1-\lambda ) R(x)},
\qquad\qquad
\psi_{\lambda\mu}(x,y)= e^{- \mu(1-\lambda) R(x) - (1-\mu) R(y)}.
\label{psiPPP}
\end{equation}
Using \eqref{psid}, the generating function of the average $\langle
d_{i,i+1}\rangle$ between the $i$-th and $(i+1)$-th points is 
\begin{equation}
\sum_{i \ge 1} \lambda^{i} \langle d_{i,i+1} \rangle 
=
\int_{-\infty}^\infty \diffd x\, \left[ e^{- (1-\lambda) R(x)}
- e^{- R(x) } \right].
\end{equation}
The probability distribution function $P_{12}(a)$ that the distance
$d_{1,2}$ is equal to $a$ and the average density $\rho(a)$ seen at
a distance $a$ from the rightmost point are given by
\begin{equation}
P_{12}(a)= \int_{-\infty}^\infty \diffd x\,r(x+a) r(x) e^{- R(x)},
\qquad\qquad
\rho(a)= \int_{-\infty}^\infty \diffd x\,r(x-a) r(x) e^{- R(x)} .
\label{P12rhoPoisson}
\end{equation}
These expressions can be understood directly from the definition of the
Poisson process or, with a little more algebra, from
(\ref{P12},\ref{rho}). One can notice that $P_{12}(a)$ and $\rho(a)$ are
given by the same expression with $a$ replaced by $-a$ and are therefore
analytic continuations of each other whenever $r(x)$ is analytic.

\subsubsection{A Poisson process with an exponential density $ e^{- \alpha x}$ }
\label{sec:PPPexp}

In the special case where the density of the Poisson process is an
exponential $r(x)= \exp({- \alpha x })$, one can simply replace $R(x)$ in
the previous expressions by $\exp(-\alpha x)/\alpha$. This gives
\begin{equation}
\begin{aligned}
\psi_{\lambda}(x)
	&= \exp \left[ -(1-\lambda ) \frac{e^{-\alpha x}}{\alpha} \right] 
	= \exp \left[ -e^{-\alpha \big(x - \frac{\ln(1-
	\lambda)}\alpha\big)}\right],
\\
\psi_{\lambda\mu}(x,y) &= \exp \left[
 - \mu (1-\lambda ) \frac{e^{-\alpha x}}\alpha
 - (1-\mu ) \frac{e^{-\alpha y}}\alpha 
\right] ,
\end{aligned}
\label{A1}
\end{equation}
so that from (\ref{psid})
\begin{equation}
\sum_{i \ge 1} \lambda^{i} \langle d_{i,i+1} \rangle = -
\frac{\ln(1 - \lambda)}\alpha,
\end{equation}
and thus \cite{BrunetDerrida.09}
\begin{equation}
\langle d_{i,i+1} \rangle = \frac 1 {\alpha \, i} .
\label{diexpo}
\end{equation}
One also has from \eqref{P12rhoPoisson}
\begin{equation}
P_{12}(a)= \alpha \, e^{-\alpha a},
 \qquad\qquad
\rho(a) = \alpha \,e^{\alpha a}.
\label{A5}
\end{equation}

\subsubsection{Decorated measures} \label{sec:Deco}

Start with a collection of points $u_i$, distributed according to some
measure~$\nu_1$ and, independently for each point $u_i$, replace it by a
realization of another measure $\nu_2$ shifted by $u_i$. We say that the
points $u_i$ are \emph{decorated} by the measure $\nu_2$ and call the
resulting measure as $\nu_1$ decorated by $\nu_2$.

We assume that $\nu_1$ and $\nu_2$ are such that the decorated measure has
a rightmost point. If the functions $\psi_\lambda(x)$,
$\psi_{\lambda\mu}(x,y)$, \ldots\ for the measure $\nu_2$ are known, the
decorated measure is characterized by functions
$\Psi_\lambda(x)$, $\Psi_{\lambda \mu}(x,y)$, \ldots\ given by
\begin{equation}
\Psi_\lambda(x)=\bigg\langle\prod_i \psi_\lambda(x-u_i)\bigg\rangle_{u_i},
\qquad\qquad 
\Psi_{\lambda \mu}(x,y)
=\bigg\langle \prod_i \psi_{\lambda\mu}(x-u_i,y-u_i)\bigg\rangle_{u_i},
\label{avgui}
\end{equation}
where the average is over all realizations $\{u_i\}$ of the
measure~$\nu_1$.
For instance, if $\nu_1$ is a Poisson process of density $r(u)$, then
\begin{equation} 
\begin{gathered}
\Psi_\lambda(x)= \prod_u \Big[1 - r(u) \diffd u + r(u) \psi_\lambda(x-u)
\diffd u\Big] = \exp \left[ \int \big[ \psi_{\lambda}(x-u) -1\big] r(u) \,
\diffd u \right],
\\
\Psi_{\lambda\mu}(x,y)=
 \exp \left[ \int \big[ \psi_{\lambda\mu}(x-u,y-u) -1\big] r(u) \,
\diffd u \right].
\end{gathered}
\label{Psi}
\end{equation}

\subsubsection{Ruelle cascades}\label{RuelleCascade}

For a decorated measure where the decoration $\nu_2$ is
a Poisson process of density $e^{-\alpha x}$, the average over the $u_i$'s
in \eqref{avgui} leads in general to complicated expressions for
$\Psi_\lambda(x)$ or $\Psi_{\lambda\mu}(x,y)$. The expressions for
$P_{12}(a)$ and $\rho(a)$ are however the same as in \eqref{A5} for the pure
Poisson process of density $e^{-\alpha x}$. In fact, all the statistical
properties of the distances between the rightmost points are the same as
those in the exponential Poisson process.

This can be understood from the following reason: decorating the
points $u_1, \ldots, u_k, \ldots$ by independent realizations of a
Poisson process of density $e^{-\alpha x}$ is equivalent to drawing a
single realization of a Poisson process of density $\sum_k e^{-\alpha
(x-u_k)}= e^{-\alpha x}\sum_k e^{\alpha u_k}$, which is just the same as
one realization of a Poisson process of density $e^{-\alpha x}$ shifted
by the random variable $\ln(\sum_k e^{\alpha u_k})/\alpha$.

The same argument applies to Ruelle cascades, which can be defined
as follows
\cite{Ruelle.87,BovierKurkova.06,BovierKurkova.07,BolthausenSznitman.98}: take an
increasing sequence of positive numbers $\alpha_1<\alpha_2<\cdots$ and
start with a Poisson process of
density $e^{-\alpha_1 x}$. At each step~$k>1$, each point in the system
is decorated by a Poisson process of density $e^{-\alpha_k x}$. At step
$k$, the measure of points in the system is simply, from the previous
argument, a Poisson process of density $e^{-\alpha_k x}$ globally shifted
by a random variable which depends on the positions of the points at step
$k-1$. Therefore, the statistics of the distances of the rightmost points
is the same as for the Poisson process of density $e^{-\alpha_k x}$.

\section{The branching Brownian motion and Fisher-KPP fronts} \label{sec:BBMFKPP}
\subsection{The Fisher-KPP equation} \label{sec:FKPP}

We are now going to see how the generating functions
(\ref{defpsi},\ref{defpsi2},\ref{defpsi3}) can be determined when the
random set of points on the line are the points generated at time~$t$ by
a branching Brownian motion.

To define the branching Brownian motion we start at time $t=0$ with
a single point at the origin. This point diffuses and branches, and its
offspring do the same. After some time $t$, a realization of the process
consists of a finite number of points located at positions $X_i(t)$ for
$i=1,2,3,\ldots $ Then, during the next time interval $\diffd t \ll 1$,
each point, independently of what the others do, moves a random
distance $X_i(t+\diffd t)- X_i(t) = \eta_i(t) \sqrt{2\diffd t}$ with
$\langle \eta_i(t) \rangle=0$ and $ \langle \eta_i(t)^2 \rangle=1 $, and,
with probability $\diffd t$, is replaced by two new points located at the
same position $X_i(t)$.

For any function $\phi$ one can define the generating function
$H_\phi(x,t)$ by
\begin{equation} 
H_\phi(x,t)
= \left\langle \prod_i \phi\big[x-X_i(t)\big]\right\rangle, \label{dual}
\end{equation} 
where the $X_i(t)$ for $i=1,\ldots, N_t$ are the positions of the $N_t$
points of the branching Brownian motion at time $t$ and $\langle \cdot
\rangle$ denotes an average over all the possible realizations. 

\begin{figure}[ht]
\centering\includegraphics{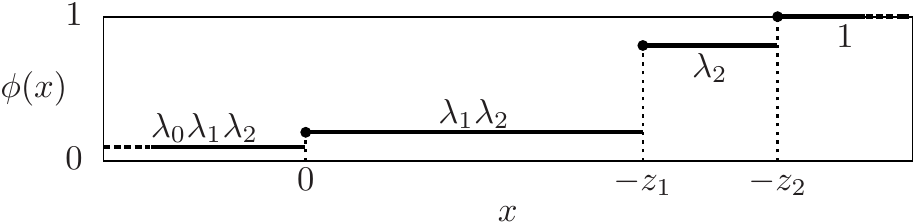}
\caption{The function \eqref{choicef} for $k=2$.}\label{figphi}
\end{figure}

By analyzing what happens during the very first time interval $\diffd t$, one can
see that the evolution of $H_\phi(x,t)$ satisfies 
\begin{equation}
H_\phi(x,t+\diffd t)=(1-\diffd t)\left\langle 
H_\phi(x-\eta\sqrt{2\diffd t},t)\right\rangle_\eta+\diffd t\,H_\phi(x,t)^2.
\label{KPP1}
\end{equation}
The first term in the right hand side represents the motion of the initial point
during the first time interval $\diffd t$ and the second term represents the
branching event which occurs with probability $\diffd t$ during this first time
interval. Taking $\diffd t$ to zero, one gets
\begin{equation}
\partial_t H_\phi = \partial_x^2
H_\phi + H_\phi^2-H_\phi,
\label{FKPPH}
\end{equation}
which is the Fisher-KPP equation \cite{Fisher.37,KPP.37,McKean.75}. (The
Fisher-KPP equation is often written as $\partial_t h = \partial_x^2
h + h -h^2$, but this is identical to \eqref{FKPPH} by the change of
variable $h=1-H_\phi$.) Because there is a single point at the origin at
time $t=0$, the initial condition is simply, from
\eqref{dual},
\begin{equation}
H_\phi(x,0)=\phi(x).
\label{initcond}
\end{equation}
The generating function (\ref{defpsi3}) at time~$t$
\begin{equation} 
\psi_{\lambda_0, \ldots, \lambda_k}(x_0, \ldots, x_k) = \left\langle
\lambda_0^{n(x_0)} \cdots \lambda_k^{n(x_k)}\right\rangle
\end{equation}
can be written, for $0> z_1 > \cdots > z_k$, as
\begin{equation}
\psi_{\lambda_0, \ldots, \lambda_k}(x, x+z_1,\ldots, x + z_k) =
 \left\langle \prod_i \phi\big[x-X_i(t)\big]\right\rangle = H_\phi(x,t), 
\label{psi-h}
\end{equation}
where the function $\phi(x)$ is given by
\begin{equation}
\phi(x) = \lambda_0^{1-\theta(x)} \lambda_1^{1-\theta(x+z_1)} \cdots
\lambda_k^{1-\theta(x+z_k) },
\label{choicef}
\end{equation}
and where $\theta(x)$ is the Heaviside step function defined by
\begin{equation}
\theta(x)=\begin{cases}1&\text{for $x\ge 0$},\\0&\text{for $x<0$}.\end{cases}
\label{teta}
\end{equation}
See figure~\ref{figphi} for the general shape of \eqref{choicef}.

With the choice (\ref{choicef}) of $\phi$, the generating function
(\ref{defpsi3}) and, therefore, all the properties of the point measure in
the branching Brownian motion at time~$t$ can be obtained as solutions of
the Fisher-KPP equation with the initial condition (\ref{initcond}).

In the special case $k=0$ and $\lambda_0=0$ of \eqref{choicef}, i.e.
for the initial condition $\phi(x)=\theta(x)$, one gets
\begin{equation}
H_\theta(x,t)=\proba(\text{There are no point at time $t$ on the right of $x$}),
\label{Hprob}
\end{equation}
and one recovers the well-known fact \cite{McKean.75,Bramson.83} that the
solution $H_\theta(x,t)$ of the Fisher-KPP equation with a step initial
condition is the cumulative distribution function of
the position of the rightmost point.

In section \ref{quantitative} we will choose $\phi=\phi_1$ and
$\phi=\phi_2$, other special cases of \eqref{choicef}, given by 
\begin{equation}
\phi_1(x)=\begin{cases}
	1 &\text{for $x\ge 0$},\\
	\lambda &\text{for $x< 0$},
 \end{cases}
\qquad\qquad
\phi_2(x)=\begin{cases}
	1 &\text{for $x \ge a$},\\
	\mu &\text{for $0 \le x< a$},\\
	\lambda\mu &\text{for $x< 0$},\\
 \end{cases}
\label{choicefsimple}
\end{equation}
to calculate the generating functions (\ref{defpsi},\ref{defpsi2}) at
time~$t$
\begin{equation}
\psi_\lambda(x) = H_{\phi_1}(x,t),\qquad\qquad
\psi_{\lambda \mu}(x,x-a) = H_{\phi_2}(x,t)
\label{phi1phi2}
\end{equation}
needed to determine the distribution $P_{12}(a)$ and the density $\rho(a)$
defined at the end of section \ref{statistics}. 

\subsection{The branching Brownian motion seen from the rightmost point} \label{BBM-frame}

The Fisher-KPP equation \eqref{FKPPH} has two homogeneous solutions:
$H_\phi=1$, which is unstable, and $H_\phi=0$, which is stable. When the
initial condition $\phi(x)$ is given by the step function $\theta(x)$, see
\eqref{teta}, the solution $H_\theta(x,t)$ of (\ref{FKPPH}) becomes
a traveling wave with the phase $H_\theta=0$ invading the phase
$H_\theta=1$ \cite{Fisher.37,KPP.37,Bramson.83}. As the front is an
extended object, one can define its position $m_t$ in several ways; for
example one could define $m_t$ as the solution of $H_\theta(m_t,t)=\alpha$
for some $0<\alpha<1$. Here it will be convenient to use the following
definition
\begin{equation}
m_t= \int\diffd x\, x \partial_x H_\theta(x,t).
 \label{defmt}
\end{equation}
One can see using (\ref{Hprob}) that $m_t$ defined by (\ref{defmt}) 
is the average position of the rightmost point.

If the initial condition (\ref{initcond}) is not a step function but is
such that $\phi(x)=1$ for all large enough $x$ and $\phi(x)$ is a
constant smaller than~1 for all large negative $x$, as in
(\ref{choicef},\ref{choicefsimple}), the solution $H_\phi(x,t)$
of (\ref{FKPPH}) becomes also a traveling wave. Its position
$m_t^{(\phi)}$ can be defined as in (\ref{defmt}) by 
\begin{equation}
m_t^{(\phi)} = \int \diffd x\,x \, \partial_x H_\phi(x,t).
\label{defmtbis}
\end{equation}
We are now going to show that the whole measure seen from the rightmost
point can be written in terms of this position $m_t^{(\phi)} $: one can
rewrite (\ref{the-frame}) as
\begin{equation}
\begin{aligned}
\chi_{\lambda_1,\ldots, \lambda_k}(z_1, \ldots, z_k) &=
\int \diffd x\, \left( \partial_x - \partial_{z_1} - \cdots
- \partial_{z_{k}} \right) \psi_{0,\lambda_1, \ldots,\lambda_k}(x,x+z_1,
\ldots, x+z_k) ,
\\
&= \psi_{0,\lambda_1, \ldots,\lambda_k}(x,x+z_1, \ldots,
x+z_k)\Big|_{x=-\infty}^{x=+\infty} 
\\
&\quad-\int \diffd x\, \left( \partial_{z_1} + \cdots+ \partial_{z_{k}}
\right) \psi_{0,\lambda_1, \ldots,\lambda_k}(x,x+z_1, \ldots, x+z_k),
 \\
&= 
1 +\left( \partial_{z_1} + \cdots + \partial_{z_{k}} \right)\int \diffd x\,
x \partial_x \psi_{0,\lambda_1, \ldots,\lambda_k}(x,x+z_1, \ldots, x+z_k). 
\end{aligned}
\label{the-frame-2-bis}
\end{equation}
Then from \eqref{psi-h} and (\ref{defmtbis})
one gets
\begin{equation}
\chi_{\lambda_1,\ldots, \lambda_k}(z_1, \ldots, z_k) = 
1 + \left( \partial_{z_1} + \cdots+ \partial_{z_{k}} \right) 
m_t^{(\phi)},
\label{the-frame-2}
\end{equation}
where $\phi$ is the function \eqref{choicef} with $\lambda_0=0$.

Therefore, \emph{with the definition (\ref{defmtbis}) of the position of
the front}, the whole information about the measure in the frame of the
rightmost point, at any time $t$, can be extracted from the $\phi$
dependence of $m_t^{(\phi)}.$

\subsection{The limiting measure and the delays}

In the long time limit, it is known \cite{Bramson.83,vanSaarloos.03} that
the traveling wave solution $H_\theta(x,t)$ of (\ref{FKPPH}), with the
initial condition (\ref{teta}), takes an asymptotic shape, $F(x)$. This
means that
\begin{equation}
H_\theta(m_t+x,t) \xrightarrow[t\to\infty]{} F(x),
\label{HF2}
\end{equation}
where $F(x)$ satisfies
\begin{equation}
F''+2F'+F^2 - F = 0, \qquad
F(-\infty)=0, \qquad
F(\infty)=1, \qquad
\int \diffd x \, x \partial_x F(x)=0.
\label{defF2}
\end{equation}
It is also known, since the work of Bramson \cite{Bramson.83}, that, in the long time limit, the
traveling wave moves at a velocity 2 and that its position (\ref{defmt}) is
given by 
\begin{equation}
m_t = 2t-\frac32\ln t + \text{Constant} + o(1). 
\label{Bramson}
\end{equation}

If the function $\phi(x)$ is not the step function but is of the form
(\ref{choicef},\ref{choicefsimple}), the solution $H_\phi(x,t)$ of
(\ref{FKPPH}) becomes also a traveling wave with the same shape $F(x)$.
This wave is centered around the position $m_t^{(\phi)}$, defined in
(\ref{defmtbis}), and one has
\begin{equation}
H_\phi(m_t^{(\phi)}+x,t) \xrightarrow[t\to\infty]{} F(x).
\label{HmphiF}
\end{equation}
For large times $m_t^{(\phi)}$ is still given by (\ref{Bramson}), but
with a different constant \cite{Bramson.83}. This means that 
\begin{equation}
m_t-m_t^{(\phi)}\xrightarrow[t\to\infty]{} f[\phi],
\label{defdelay}
\end{equation}
where $f[\phi]$ is the long time delay in the position of the front due
to the modified initial condition, as compared to a front starting with a
step function. Taken together, \eqref{HmphiF} and \eqref{defdelay} give
\begin{equation}
H_\phi(m_t+x,t)  \xrightarrow[t\to\infty]{} F\big(x+f[\phi]\big).
\label{delay}
\end{equation}
Using \eqref{psi-h}, this becomes 
\begin{equation}
\psi_{\lambda_0, \ldots, \lambda_k}(m_t+x, m_t+x+z_1,\ldots, m_t+x + z_k)
 \xrightarrow[t\to\infty]{} F\big(x+f[\phi]\big),
\label{stat1}
\end{equation}
which shows that the measure of $\{X_1(t)-m_t, X_2(t)-m_t, \ldots\}$ (the
rightmost points in the branching Brownian motion seen from the $m_t$
frame) does converge when $t\to\infty$ to a limiting point measure
characterized by the functions $F(x)$ and $f[\phi]$.

The measure of $\{X_2(t)-X_1(t), X_3(t)-X_1(t), \ldots\}$ (the rightmost
points in the branching Brownian motion seen from the $X_1(t)$ frame) also
has a well-defined limit when $t\to\infty$. Indeed, using
(\ref{the-frame-2}) and \eqref{defdelay}, one gets
\begin{equation}
\chi_{\lambda_1,\ldots, \lambda_k}(z_1, \ldots, z_k) \xrightarrow[t\to\infty]{} 
1 - \left( \partial_{z_1} + \cdots +\partial_{z_{k}} \right) 
f[\phi].
\label{the-frame-3}
\end{equation}
Therefore, in the long time limit, all the information on the
distribution of the rightmost points seen from $X_1(t)$ is
contained in the $\phi$ dependence of the delay $f[\phi]$.

Note that, in contrast to (\ref{the-frame-2}) which requires the
position to be defined by (\ref{defmtbis}), the delay $f[\phi]$ in
\eqref{stat1} or (\ref{the-frame-3}) depends only on $\phi$: it would not
change if we had chosen another definition of the front position.

\subsection{The superposability of branching Brownian motions}\label{superpos1}

Let us now consider $M$ independent branching Brownian motions starting at
$t=0$ at positions $u_1, \ldots, u_M$. Following the same argument as in
section \ref{sec:FKPP}, the generating function (\ref{defpsi3}) of the
union of the points at time $t$ of these $M$ branching Brownian motions is
given by the following generalization of~(\ref{psi-h})
\begin{equation}
\psi_{\lambda_0, \ldots , \lambda_k}(x, x+z_1,\ldots, x + z_k) =
\prod_{\alpha=1}^M H_\phi(x-u_\alpha,t),
\end{equation}
where $H_\phi(x,t)$ is the same solution of (\ref{FKPPH}) with the initial
condition (\ref{choicef}) as in the case of a single branching Brownian
motion starting at the origin. In the long time limit, using \eqref{delay},
\begin{equation}
\psi_{\lambda_0, \ldots, \lambda_k}(m_t+x, m_t+x+z_1,\ldots, m_t+x + z_k)
\xrightarrow[t\to\infty]{} 
\prod_{\alpha=1}^M F(x+ f[\phi]-u_\alpha) .
\label{stat2}
\end{equation}
This means that here again, there is a limiting measure when $t\to\infty$
for the rightmost points in the $m_t$ frame. This measure is not the same
as before (when one starts with a single point at the origin), as can be
seen by comparing \eqref{stat2} and \eqref{stat1}. In particular, the
distribution of the rightmost point is different.

In the frame of the rightmost point, however, one can see using 
\eqref{stat2} and (\ref{the-frame-2-bis}) that
\begin{equation}
\chi_{\lambda_1,\ldots, \lambda_k}(z_1, \ldots, z_k)\xrightarrow[t\to\infty]{} 
1 - \left( \partial_{z_1} + \cdots + \partial_{z_{k}} \right)
f[\phi],
\end{equation}
as in \eqref{the-frame-3}.

It is remarkable that the generating function $\chi$ does depend neither on
the number $M$ of starting points nor on their positions $u_\alpha$. The
picture which emerges is that if we superpose the rightmost points of
several branching Brownian motions, starting at arbitrary positions, the
limiting measure in the frame of the rightmost point is, when $t \to
\infty$, the same as for a single branching Brownian motion.

We will say that, in the long time limit, the measure of the distances
between the rightmost points in a branching Brownian motion becomes
\emph{superposable}: the union of two (or more) realizations of the process
(even moved by arbitrary translations $u_\alpha$) leads to the same measure
in the frame of the rightmost point as for a single branching Brownian
motion.

As a remark, it is easy to check, that the Poisson process with an
exponential density $r(x)= e^{- \alpha x}$, see section~\ref{sec:PPPexp},
is an example of a superposable measure: the superposition of $M$ such
Poisson processes translated by arbitrary amounts $u_1, \ldots, u_M$ is
identical to a single Poisson process with an exponential distribution
translated by $ \alpha^{-1} \ln ( e^{\alpha u_1} + \cdots + e^{\alpha u_M})
$. One can also check that, for the same reason, all the decorated measures
of section~\ref{sec:Deco} are superposable when~$\nu_1$ is a Poisson
process with an exponential density.

In section \ref{statmeas}, we will state a stronger version of the
superposability property of the branching Brownian motion.

\section{Some quantitative properties of the branching Brownian motion seen from the rightmost point}
\label{quantitative}

In this section we obtain, by integrating numerically the equation
(\ref{FKPPH}) with the appropriate initial condition, some statistical
properties of the limiting measure seen from the rightmost point.

\subsection{Average distances between consecutive points}

The analytic calculation of the delay $f[\phi]$ is in general not easy.
For $\phi=\phi_1$ given by (\ref{choicefsimple}), however, it was possible to
show \cite{BrunetDerrida.09} that when $1- \lambda \ll1$, the delay is given by
\begin{equation}
f[\phi_1] \simeq - \ln(1-\lambda) - \ln [ - \ln(1-\lambda)] + {\cal
O}(1),
\label{fphi1asympt}
\end{equation}
and, from this, one could deduce that, in the long time limit,
the average of the distance $d_{i,i+1}$ between the $i$-th and the $(i+1)$-th
rightmost points is given for large $i$ by
\begin{equation}
\langle d_{i,i+1}\rangle\simeq \frac1i-\frac1{i\ln i}.
\label{asympdi}
\end{equation}
In \cite{BrunetDerrida.09}, the numerical values of the distances between
the rightmost points were also obtained by integrating the Fisher-KPP equation
with the initial condition $\phi_1$ in (\ref{choicefsimple}) and by using
(\ref{psid}) (in practice we integrated numerically the equations satisfied
by the coefficients of the expansion of $\psi_\lambda(x)$ in powers of
$\lambda)$. It was found that
\begin{equation}\begin{aligned}
 \langle d_{1,2}\rangle&\simeq0.496,
&\langle d_{2,3}\rangle&\simeq0.303,
&\langle d_{3,4}\rangle&\simeq0.219,\\
 \langle d_{4,5}\rangle&\simeq0.172,
&\langle d_{5,6}\rangle&\simeq0.142,
&\langle d_{6,7}\rangle&\simeq0.121.
\label{valuesdn}
\end{aligned}\end{equation}
The results (\ref{asympdi},\ref{valuesdn}) gave evidence that the
distances between the rightmost points of the branching Brownian motion
were different from those of a Poisson process with an exponential
density (\ref{diexpo}).

\subsection{Distribution of the distance between the two rightmost points}
\label{sec:distances}

According to (\ref{P12}), to obtain the distribution $P_{12}(a)$ of the
distance between the two rightmost points, one needs to calculate
$\psi_{0\mu}(x,x-a)$ to first order in $\mu$. We first remark that at
time~$t$ for $a>0$, $\psi_{00}(x,x-a)=\proba[ n(x)=0 \AND n(x-a)=0 ]=
\proba[n(x-a)=0]=H_\theta(x-a,t)$ where $H_\theta(x,t)$ is the standard
Fisher-KPP front with the step initial condition (it is also easy to see
from the definition \eqref{choicefsimple} of $\phi_2$.) Then writing at time $t$ that
\begin{equation}
\psi_{0\mu}(x,x-a) = H_\theta(x-a,t)+\mu R_a(x-a,t) + {\cal O}(\mu^2)
\label{expnpsiR}
\end{equation}
is solution of the Fisher-KPP equation, and using the initial
condition $\phi_2$ in \eqref{choicefsimple}, one gets
\begin{equation}
\partial_t R_a = \partial_x^2 R_a -R_a +2 H_\theta R_a;
\qquad
R_a(x,0) = \begin{cases} 
		1 &\text{for $-a\le x<0$},\\
		0 & \text{otherwise}.
	 \end{cases}
\label{eqRa}
\end{equation}
Then, from \eqref{P12} one gets
\begin{equation}
P_{12}(a) = -\partial_a^2\int \diffd x \, R_a(x,t).
\label{RaP12}
\end{equation}

\begin{figure}[ht]
\centering
\includegraphics[height=.4\textwidth]{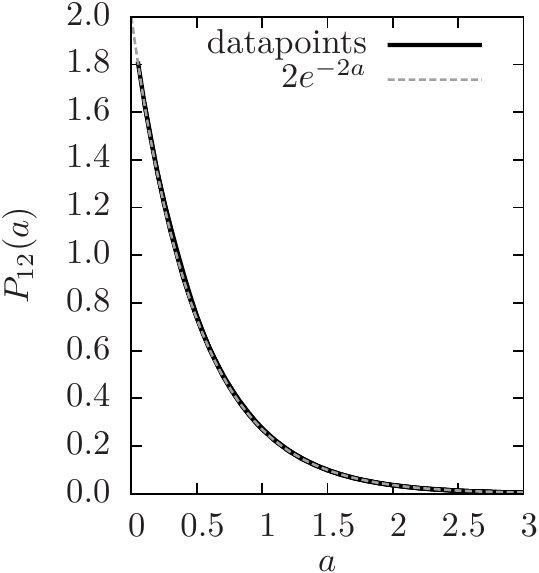}\quad
\includegraphics[height=.4\textwidth]{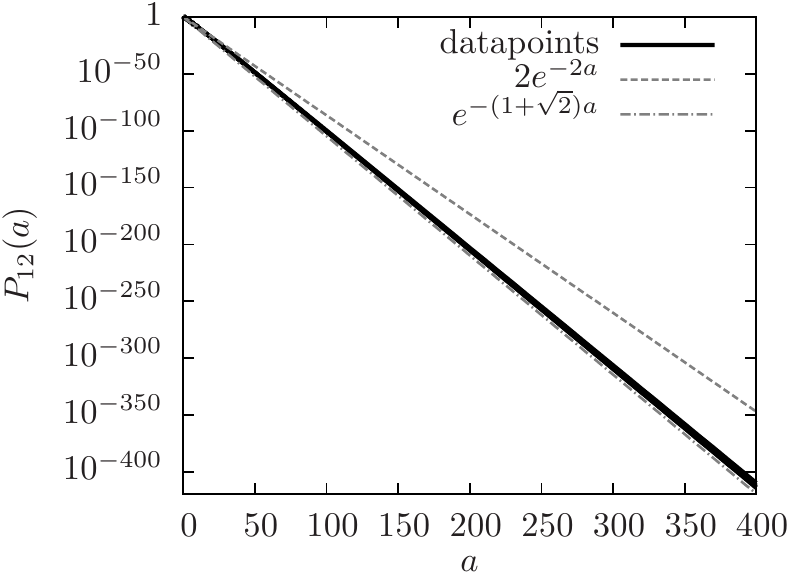}
\caption{The density of probability $P_{12}(a)$ of observing a distance $a$
between the two rightmost points in the $t\to\infty$ limit, as a function
of $a$. For $a$ small (left part), the distribution is very close to
$2e^{-2a}$. For larger values of $a$, one observes a faster exponential
decay of order $e^{-(1+\sqrt2)a}$.}
\label{fig:distrib_d}
\end{figure}

Figure~\ref{fig:distrib_d} shows our numerical result for the distribution
$P_{12}(a)$ of the distance between the two rightmost points in the long
time limit. More details on our numerical procedure is given in
appendix~\ref{sec:numeric}. 

We see that $P_{12}(a)$ is very close to $2 e^{-2a}$ for the values of $a$
which have a significant probability of occurring. This is of course
consistent with an average distance (\ref{valuesdn}) close to $1/2$. For
large $a$ (events with a small probability), however, the exponential
decay is faster. We now present a simple argument leading to the
following prediction, which is consistent with our numerical data,
\begin{equation}
P_{12}(a) \sim e^{-(1+\sqrt2) a}\qquad\text{for large $a$}.
\label{decayratedist2}
\end{equation}
In the long time limit, the right frontier of the branching Brownian motion
moves at velocity $v=2$. Let us assume that a large distance $a$ between
the two rightmost points is produced by the following scenario: 
by a rare fluctuation, the rightmost point escapes and, without branching,
goes significantly ahead while the rest of the points go on as usual,
forming a frontier moving at velocity~$v=2$. Such an event leads to the
distance $a$ between the two rightmost points if, during a time $\tau$, the
rightmost point moves (by diffusion alone) by a distance $a+ 2 \tau$
without branching. The probability of such a scenario is
\begin{equation}
\proba(\text{$X_1-X_2\simeq a$ after an escape time $ \tau$})\sim
\exp\left[{-\frac{(a+2\tau)^2}{4\tau}}\right]\times e^{-\tau}.
\label{prob1}
\end{equation}
The first term is the probability of diffusing over a distance $a+2\tau$
during time $\tau$, and the second term is the probability of not
branching. The probability to observe a large distance $a\gg1$ is then
dominated by the events with $\tau$ chosen to maximize 
\eqref{prob1}, that is 
\begin{equation}
\tau_\text{optimal}=\frac{a}{2\sqrt2},
\end{equation}
and this leads to (\ref{decayratedist2}) in good agreement with the
numerical data of figure \ref{fig:distrib_d}.

There is a remarkable relation between the decay rate in
(\ref{decayratedist2}) and the shape of the traveling wave solution of
(\ref{defF2}). Around the \emph{stable} region $F=0$, the equation
\eqref{defF2} can be linearized and one has
\begin{equation}
F(z)\simeq \text{Cste} \times e^{r z} \text{ for
$z\to-\infty$,}\qquad\text{with}\quad
r^2+2r-1=0.
\end{equation}
We emphasize that this is a linear analysis of the \emph{stable} region, which
is usually uninteresting (in contrast to the \emph{unstable} region which
determines the velocity). The solutions for $r$ are
\begin{equation}
r=-1\pm\sqrt{2}.
\end{equation}
$r=-1+\sqrt{2}$ is the correct root as it is the only positive solution and
$F(-\infty)$ has to vanish. The other solution $r=-1-\sqrt2$ (the wrong
root) coincides (up to the sign) with the decay rate of the distribution
$P_{12}(a)$ for the distance $a$ between the two rightmost points
(\ref{decayratedist2}).

As explained in appendix~\ref{sec:largedeviation}, this coincidence exists
in a broad class of branching processes: each variant of the branching
Brownian motion is linked to a variant of the Fisher-KPP equation, and the
wrong root in the linear analysis of the stable region always gives the
asymptotic decay rate of $P_{12}(a)$.

\subsection{Average density seen from the rightmost point} \label{sec:density}

To obtain the average density of points at a distance $a$ on the left of
the rightmost point, one needs, according to~\eqref{rho}, to calculate
$\psi_{0\mu}(x,x-a)$ for $\mu$ close to~1. As in
section~\ref{sec:distances}, we first remark, from the
definition~\eqref{defpsi2}, that
$\psi_{01}(x,x-a)=\proba[n(x)=0]=H_\theta(x,t)$ is the standard Fisher-KPP
front with the step initial condition. Then, writing at time~$t$ that
\begin{equation}
\psi_{0\mu}(x,x-a)=H_\theta(x,t) -(1-\mu)\tilde R_a(x,t)+{\cal
O}[(1-\mu)^2]
\label{expnpsiRtilde}
\end{equation}
is solution of the Fisher-KPP equation, and using the initial
condition $\phi_2$ in \eqref{choicefsimple}, one gets
\begin{equation}
\partial_t \tilde R_a = \partial_x^2 \tilde R_a -\tilde R_a +2 H_\theta
\tilde R_a;
\qquad
\tilde R_a(x,0) = \begin{cases} 
		1 &\text{for $0\le x<a$},\\
		0 & \text{otherwise}.
	 \end{cases}
\label{eqRatilde}
\end{equation}
It is the same equation as for $R_a$ in \eqref{eqRa}, but with a different
initial condition. Then, from \eqref{rho} one gets
\begin{equation}
\rho(a) = \partial_a^2\int \diffd x \, \tilde R_a(x,t).
\label{Rarho}
\end{equation}

One can notice the great similarity between the expressions for the average
density $\rho(a)$ of points at a distance~$a$ from the rightmost
(\ref{eqRatilde},\ref{Rarho}) and the probability distribution $P_{12}(a)$
for the distance between the two rightmost points (\ref{eqRa},\ref{RaP12}):
one goes from one to the other by simple changes of signs, as in the
example of a Poisson process \eqref{P12rhoPoisson}.

\begin{figure}[ht]
\centering
\includegraphics[width=.5\textwidth]{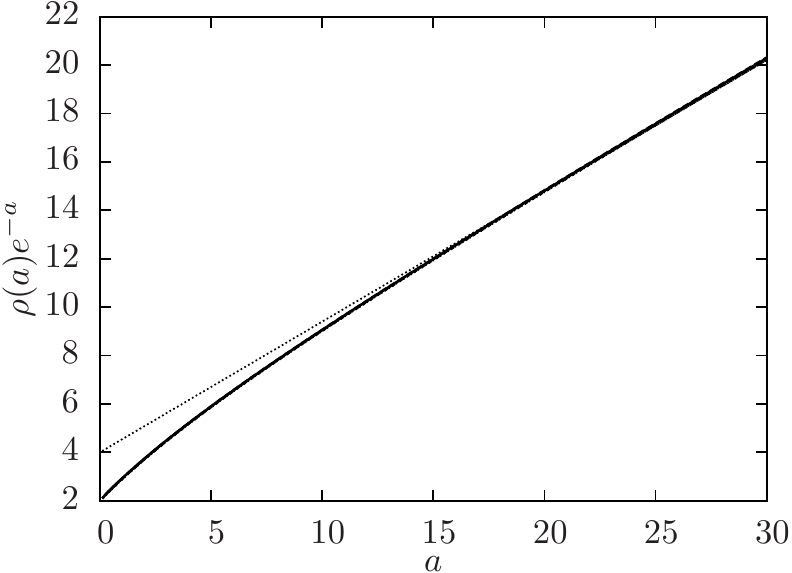}
\caption{The average density $\rho(a)$ of points at a distance $a$ of the rightmost
in the long time limit grows like $ a e^{a}$. When the data is multiplied
by $e^{-a}$, as shown in the figure, the linear prefactor is clearly
visible.}
\label{fig:density}
\end{figure}

Figure~\ref{fig:density} presents our numerical results for~$\rho(a)$ in
the long time limit. We see that $\rho(a)$ increases as 
\begin{equation}
\rho(a)\simeq \text{Cste} \times a e^a\qquad\text{for large $a$}.
\label{aexpa}
\end{equation}
Note that a Poisson process with such a density would lead to asymptotic
distances between points given by \eqref{asympdi}. The branching Brownian
motion is however not a Poisson process as the points are correlated, at
least near the tip.

\section{Generalizations to other branching processes} \label{sec:disc}

All the results of sections~\ref{sec:BBMFKPP} and \ref{quantitative} can be
generalized to other branching processes on the line where points move and
branch independently of the positions and of the motions of the other
points. In such systems, the function $H_\phi(x,t)$ defined in
\eqref{dual} is also solution of an equation similar to the Fisher-KPP
equation~\eqref{FKPPH}. Here are four examples:
\begin{itemize}
\item[A)] The points perform Brownian motions and branch as before, but at each
branching event there is a probability $p$ to branch into three points
and $1-p$ to branch into~two. Then $H_\phi(x,t)$ evolves according to
\begin{equation}
\partial_t H_\phi=\partial_x^2 H_\phi + p H_\phi^3 + (1-p) H_\phi^2
-H_\phi.
\label{ex1}
\end{equation}
\item[B)] Time is discrete with steps of duration $\delta$; at each time
step, a point at position $x$ branches into two points at positions
$x+\epsilon_1$ and $x+\epsilon_2$, where the $\epsilon_i$ take
independent random values distributed according to some given
$\rho(\epsilon)$. The evolution of $H_\phi(x,t)$ is then given by
\begin{equation}
H_\phi(x,t+\delta)=\left[\int \diffd\epsilon\, \rho(\epsilon)H_\phi(x-\epsilon,t)\right]^2.
\label{ex3}
\end{equation}
In this example, the positions of the points can be thought of as the possible
energies of a directed polymer on a Caley tree with independent random
energies $\epsilon$ on the edges of the tree
\cite{DerridaSpohn.88,MonthusGarel.07}.
\item[C)] Time is continuous but space is discrete with steps~1; during
$\diffd t$, each point at position~$x$ has a probability $\diffd t$ of
being removed and replaced by two points at position~$x+1$. The equation
satisfied by $H_\phi(x,t)$ is
\begin{equation}
\partial_t H_\phi(x,t)= H_\phi(x-1,t)^2 -H_\phi(x,t).
\label{ex4}
\end{equation}
This example is relevant to the theory of binary search trees
\cite{Roberts.10,MajumdarKrapivsky.02,MajumdarDeanKrapivsky.05}.
\item[D)] Time and space are discrete with steps $s$ for space and
$\delta$ for time;
in a given time step, a point at position $x$
has a probability $\delta$ of branching into two points at position $x$,
a probability $\delta/s^2$ of jumping to the left, $\delta/s^2$ of jumping
to the right, and $1-\delta-2\delta/s^2$ of doing nothing. Then:
\begin{equation}
H_\phi(x,t+\delta) = H_\phi(x,t)+\delta\left[\frac{H_\phi(x-s,t)
+ H_\phi(x+s,t)-2H_\phi(x,t)}{s^2} -H_\phi(x,t) + H_\phi(x,t)^2 \right],
\label{ex5}
\end{equation}
which is of course a discretized version of the original Fisher-KPP equation.
\eqref{ex5} is actually the equation we used in our numerical simulations,
see appendix~\ref{sec:numeric}.
\end{itemize}
In all cases, these equations have $H_\phi=1$ as an unstable fixed point,
and $H_\phi=0$ as a stable fixed point. For initial conditions
$H_\phi(x,0)=\phi(x)$ of the type
(\ref{choicef},\ref{teta},\ref{choicefsimple}), the function $H_\phi(x,t)$
develops into a traveling wave moving at a specific velocity $v^*$.
We recall briefly the procedure to determine the asymptotic velocity~$v^*$ of the
front (which is also, through~\eqref{Hprob}, the velocity of the rightmost
point in the branching process). One looks for traveling wave solutions
moving at velocity~$v$ of
the form $H_\phi(x,t)=F(x-vt)$ and solve the linearized equation around the
unstable fixed point by writing $1-F(x)\simeq \epsilon e^{-\gamma x}$.
This leads to a relation between $\gamma$ and $v$, and the minimal value
$v^*$ of $v$ reached at some $\gamma^*$ is the velocity selected by the
front \cite{vanSaarloos.03}. (We only consider here cases where the
function $v(\gamma)$ has a minimum.) For our four examples
\begin{equation}
\begin{aligned}
\text{A)}\quad& v=\gamma+\frac{1+p}\gamma;\ v^*=2\sqrt{1+p},
&
\text{B)}\quad&
v=\frac1{\gamma\delta}\ln\left[2\int\diffd\epsilon\,\rho(\epsilon)e^{\gamma\epsilon}\right],
\\\text{C)}\quad&
v= \frac{2e^\gamma-1}\gamma;\ v^*\simeq 4.311,
&\text{D)}\quad&
v=\frac1{\gamma\delta}\ln\left[1+2\delta\frac{\cosh(\gamma
s)-1}{s^2}+\delta\right].
\end{aligned}
\label{vgammas}
\end{equation}

Once the equation for $H_\phi$ of a particular branching process is
written, one has access to all the generating functions
$\psi_{\lambda}(x)$, $\psi_{\lambda\mu}(x,x-a)$, etc., see
(\ref{defpsi},\ref{defpsi2},\ref{defpsi3}), by choosing the appropriate
initial conditions (\ref{choicef},\ref{choicefsimple}) for the front
equation. The whole measure in the frame of the rightmost point is then
obtained from \eqref{the-frame-2} at any finite time~$t$. Note that to
prove the existence of a long time limit to the point measure in this
frame for a specific branching process, one would need a version of
Bramson's result \eqref{Bramson} for this process which is, to our
knowledge, not known in the general case. 

It is natural to ask which properties of the branching Brownian motion
can be extended to other branching processes. If the measure for the
distances between the rightmost points has a long time limit, then the
arguments of section~\ref{superpos1} can be 
easily generalized and one can show that it is superposable.
We have checked that the analytical argument
\cite{BrunetDerrida.09} leading to the asymptotic expression
\eqref{asympdi} for the average distances $\langle d_{i,i+1}\rangle$ at
large times can be extended in
case B) for a large class of densities $\rho(\epsilon)$ and yields
\begin{equation}
\langle d_{i,i+1}\rangle \simeq
\frac1{\gamma^*}\left(\frac1i-\frac1{i\ln i}\right)\qquad\text{for large
$i$}.
\label{genasymptdi}
\end{equation}
We have also checked numerically on examples C) and D) that
the density at a distance $a$ of the rightmost point is, as in \eqref{aexpa},
\begin{equation}
\rho(a)\simeq \text{Cste} \times a e^{\gamma^* a}\qquad\text{for large $a$}.
\label{genaexpa}
\end{equation}
For the tail of the distribution $P_{12}(a)$ of the distance $a$ between
the two rightmost points, we discussed a scenario, at the end of
section~\ref{sec:distances}, which can be generalized (see
appendix~\ref{sec:largedeviation}) to calculate the exponential decay
$P_{12}(a)$ for more general branching processes. This scenario, however,
can only hold if points can move without branching, as in our examples~A)
and~D); for instance, in  example~A), it predicts an exponential with a
decay rate equal to $\sqrt{1+p}+\sqrt{2+p}$. In examples B) and C), the
points branch whenever they move and the
tail of $P_{12}(a)$ is in general not an exponential.

Note that special care should be taken if the points are located  on a
discrete lattice, as in cases~C), D) and possibly~B): quantities such as
$P_{12}(a)$, see (\ref{P12}), become probabilities rather than densities
of probability and quantities such as $\rho(a)$, see (\ref{rho}), become
average numbers rather than average densities, and all the formulas in the
previous sections need to be adapted: integrals become discrete sums,
derivatives become finite differences, etc. If one interprets $n(x)$ as
the number of points \emph{strictly} on the right of $x$, then the
generating functions $\psi_\lambda$, $\psi_{\lambda\mu}$, \ldots\ are
still related to $H_\phi$ as in (\ref{psi-h},\ref{phi1phi2}) with the
choices (\ref{choicef},\ref{choicefsimple}) for the initial condition
$\phi$. Then, for instance, one can show easily that \eqref{psid} becomes
\begin{gather}
\sum_x x
\big[\psi_0(x)-\psi_0(x-s)-\psi_\lambda(x)+\psi_\lambda(x-s)\big]
=\sum_{i\ge1} \langle d_{i,i+1}\rangle \lambda^i,
\end{gather}
where $s$ is the lattice spacing. For these systems on the lattice, there
are new properties that can be investigated. As an example, if $N$ is the
number of points on the rightmost occupied site, then it is easy to check
that
\begin{equation}
\big\langle\mu^N\big\rangle=1+\sum_x\big[\psi_{0\mu}(x,x-s)-H_\theta(x,t)\big].
\end{equation}
The whole distribution of
$N$ can then be determined by numerical integration. In the case of our
example~C), the number~$N$ corresponds to the number of leaves at the
deepest level in a binary search tree~\cite{Roberts.10} and
we found numerically that, at large times, $\proba(N=2)\simeq 0.50$,
$\proba(N=4)\simeq0.23$, $\proba(N=6)\simeq0.11$,
$\proba(N=8)\simeq0.06$, etc.

\section{Large time measure in the frame of Lalley and Sellke}\label{statmeas}

\begin{figure}[!ht]
\centering
\includegraphics[width=.48\textwidth]{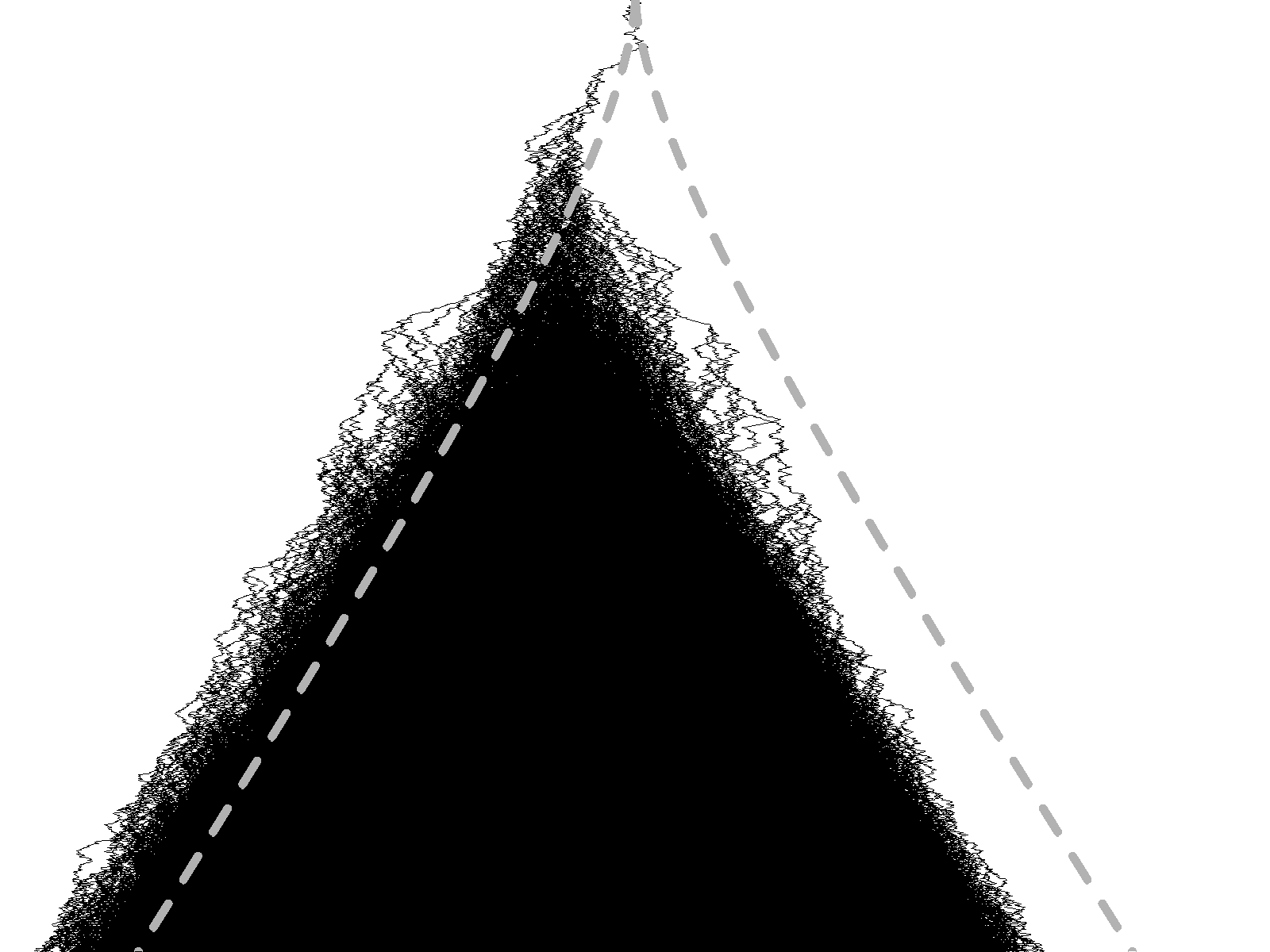}\quad
\includegraphics[width=.48\textwidth]{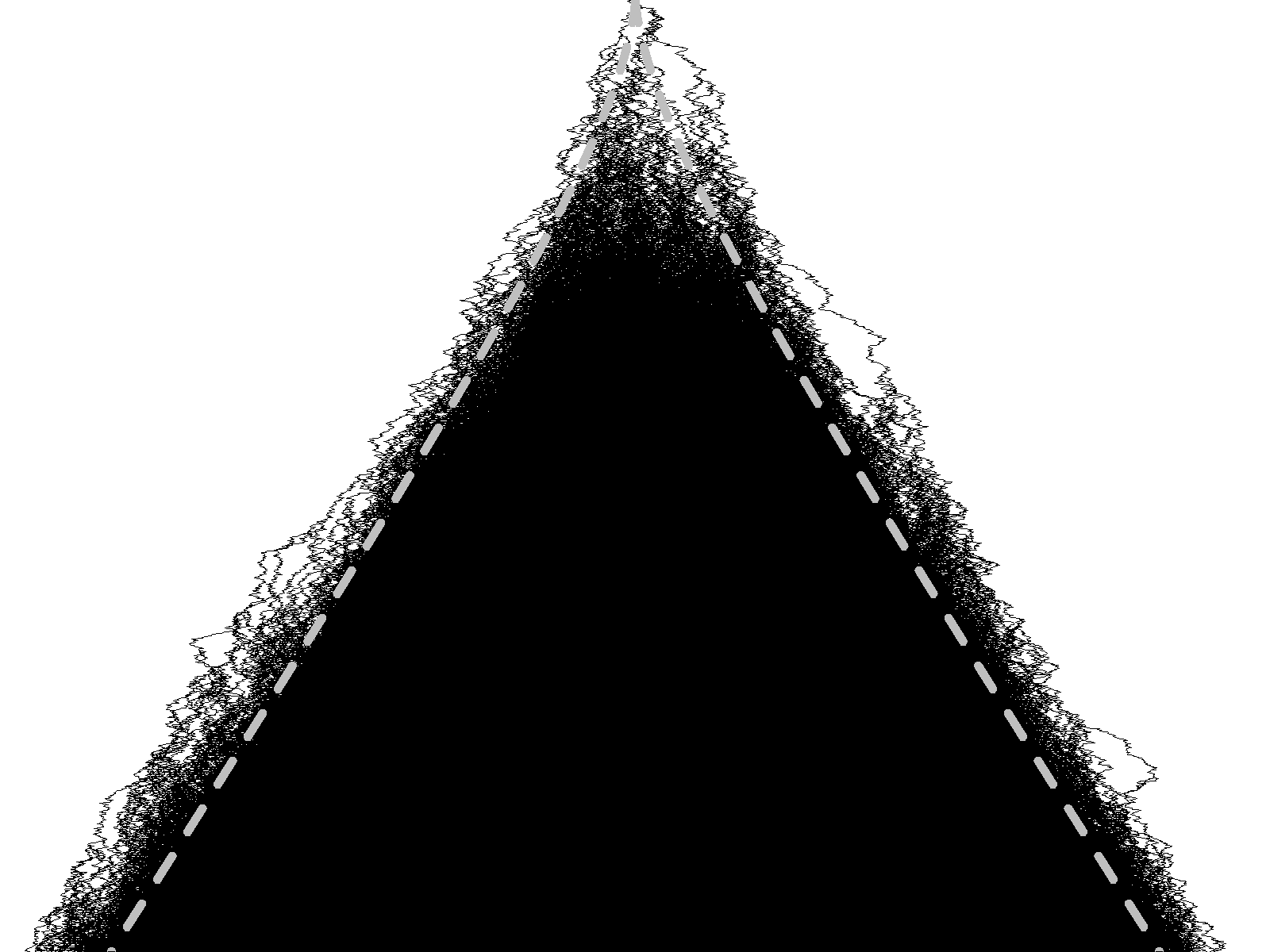}
\caption{Two realizations of the branching Brownian motion up to time
$t=20$. The horizontal direction represents space, and time increases
downwards. The dotted gray lines are $m_t$ and $-m_t$, the average
positions of the rightmost and leftmost points, as measured from
equation~\eqref{defmt}.}
\label{fig2BBM}
\end{figure}

We recall from the results of section~\ref{sec:FKPP}, see \eqref{HF2} and
\eqref{Hprob}, that the distribution of the rightmost point is given in the
long time limit by
\begin{equation}
\lim_{t\to\infty}
\proba[X_1(t)<m_t+x]=F(x),
\label{nerg}
\end{equation}
where $m_t$ is the average position of the rightmost point and is
asymptotically given by \eqref{Bramson} and where $F$ is defined in
\eqref{defF2}. The property \eqref{nerg} is however non-ergodic, as shown
by Lalley and Sellke \cite{LalleySellke.87} in the sense that for a single
realization of the branching Brownian motion
\begin{equation}
\big[\text{fraction of the time that $X_1(t)<m_t+x$}\big]\ne F(x).
\label{nonergo}
\end{equation}
In fact, the left hand side of \eqref{nonergo} is not self-averaging and
depends on the realization. This is illustrated in figure~\ref{fig2BBM}:
for the realization on the left, there were few branching events at early
times and the first points wandered to the left, leading at larger times
to an asymmetric picture. For the realization on the right of
figure~\ref{fig2BBM}, there were many branching events early. For the
right realization, the rightmost point is almost always on the right of
$m_t$ while it is almost always on the left of $m_t$ in the left
realization.

Visually, these strong memory effects of the early stages of the branching
Brownian motion do not seem to decay with time, and it looks like the
fluctuating right frontier of the system settles at some random fixed
distance $C$ from~$m_t$:
\begin{equation}
X_1(t)=m_t + C + \eta_1(t) \qquad\text{for large $t$},
\label{whatIneed}
\end{equation}
where $C$ would depend on the realization but not the time, and where
$\eta_1(t)$ would be a time-dependent random number centered around zero.
A natural question is whether it is possible to define $C$ for each
realization in such a way that the distribution of $\eta_1(t)$ becomes in
the long time limit independent of $C$ and $t$, the idea being that the
branching Brownian motion at long times seen from $m_t+C$ would ``look the
same'' for any realization, whatever is the value of $C$.

A related question was addressed by Lalley and Sellke \cite{LalleySellke.87} in
the following way: for each realization of the branching Brownian motion,
define $Z$ as
\begin{equation}
Z=\lim_{t\to\infty} Z_t\qquad\text{where }
Z_t=\sum_i [2t-X_i(t)]e^{X_i(t)-2t}.
\label{defZ}
\end{equation}
(The sum is over all the points $X_i(t)$ present at time~$t$.) As shown
in~\cite{LalleySellke.87}, $Z_t$ has a limit $Z$ for almost every
realization; that limit is finite and positive. Lalley and Sellke
prove then a limit theorem for the frontier of the branching Brownian
motion which we interpret as follows:
\begin{equation}
\lim_{t\to\infty}
\proba\big(X_1(t)<m_{t}+x \big| Z \big)=\exp\big(- A Z e^{-x}\big)
=\exp\big(- e^{-[x-\ln(A Z)]}\big),
\label{LSint}
\end{equation}
where $A$ is a constant related to the large $x$ behavior of $F(x)$,
see~\eqref{asymptF}. In words, if one considers only the realizations of
the branching Brownian motion with a given value of $Z$, then the large
time distribution of the rightmost point is given by a Gumble located
around $m_t+\ln(A Z)$. To make the link with~\eqref{whatIneed},
\begin{equation}
C=\ln(A Z),\qquad \proba(\eta_1<x)=\exp\big(-e^{-x}\big).
\end{equation}
In appendix~\ref{apxLS}, we present the precise theorem stated by Lalley
and Sellke and we argue that \eqref{LSint} should be equivalent to their
result.

A natural extension to Lalley's and Sellke's result is to write for all
the points~$i$
\begin{equation}
X_i(t) = m_t + \ln(AZ) + \eta_i(t),
\label{etai}
\end{equation}
as in \eqref{whatIneed}, and ask whether the joint distribution of
$\eta_1(t)$, $\eta_2(t)$, $\eta_3(t)$, \ldots\ reaches a
long time limit which is independent of the value of~$Z$.
We show in appendix~\ref{apxLS} that Lalley's and Sellke's result can be
extended to all the generating functions $H_\phi$. Our interpretation of
this extension is
\begin{equation}
\lim_{t\to\infty}
\Big\langle \prod_i \phi\big(m_t+x-X_i(t)\big) \Big| Z \Big\rangle
=
\exp\big({-A Z e^{-x-f[\phi]}}\big)
=\exp\big(- e^{-[x-\ln(AZ)+f[\phi]]}\big),
\label{LS2int}
\end{equation}
where the delay function $f[\phi]$ is the same as in \eqref{delay}. By
choosing $\phi=\theta$ (the step function), \eqref{LS2int} reduces to
\eqref{LSint}. By choosing $\phi$ as in \eqref{choicef}, one sees from
\eqref{LS2int} that the distribution of points at the right of the
branching Brownian motion conditioned by $Z$ reaches a long time limit
where $Z$ only appears through the global shift $\ln(AZ)$. This means that
at large times, the distribution of the rightmost points in a branching
Brownian motion has a well defined measure \emph{independent of $Z$}
located around $m_t+\ln(AZ)$.

As an example, if one chooses the function~$\phi_1$ defined by
\eqref{choicefsimple}, one can easily show from \eqref{LS2int}
and~\eqref{fphi1asympt} that, in the $m_t+\ln(AZ)$ frame,
the average density of points at any position is infinite in the long
time limit.

\subsection{Superposability property}

If one considers two branching Brownian motions $a$ and $b$ starting at
arbitrary positions, then the points in $a$ at  large time will be
characterized by a random value $Z^{(a)}$ and a realization of the point
measure described by \eqref{LS2int}; idem for the points in $b$. If one
considers the union of these two branching Brownian motions, one gets
from~\eqref{LS2int}
\begin{equation}
\lim_{t\to\infty}
\Big\langle \prod_i \phi\big(m_t+x-X_i(t)\big) \Big| Z^{(a)},Z^{(b)} \Big\rangle
=
\exp\big({-A Z e^{-x-f[\phi]}}\big)
=\exp\big(- e^{-[x-\ln(AZ)+f[\phi]]}\big),
\end{equation}
with $Z=Z^{(a)}+Z^{(b)}$. This means that the point measure reached in
the long time limit in the $m_t+\ln(AZ)$ frame is the same whether one
started initially with one, two or, by extension, any finite number of
initial points at arbitrary positions on the line. What does depend on
the initial number of points is only the law of the random number~$Z$,
not the positions around $m_t+\ln(AZ)$. This is to be related to the
discussion in section~\ref{superpos1}, where we showed that, in the long
time limit, the measure seen from $m_t$ depends on the initial number of
points while the measure seen from $X_1(t)$ does not. 

Furthermore, the large time measure of the points in the $m_t+\ln(AZ)$ frame has the
following property:

\begin{equation}
\parbox{.67\textwidth}{
Starting with two realizations $\{\eta_i^{(a)}\}$ and
$\{\eta_i^{(b)}\}$,
then for any pair of real numbers $\alpha$ and $\beta$,
the ensemble of points $\{\eta_i^{(a)}+\alpha\}
\cup \{\eta_i^{(b)}+\beta\}$ is another realization of the same measure
shifted by $\ln(e^\alpha+e^\beta)$.}
\label{supprop}
\end{equation}
(Think of $\{\eta_i^{(a)}\}$ as the offspring of $a$ in the
$m_t+\ln[AZ^{(a)}]$ frame and of $\alpha$ as $\ln Z^{(a)}$; idem
for~$b$. The shifts $\alpha$ and $\beta$ are arbitrary because $Z^{(a)}$
and $Z^{(b)}$ are unbounded independent random numbers.)

The property~\eqref{supprop} of the point measure in the $m_t+\ln(AZ)$ frame is a
stronger version of the superposability property discussed in
section~\ref{superpos1}: clearly, it implies that the distribution of distances between
the rightmost points is invariant by superposition, but it gives more
information on the measure as it encompasses the position of the
rightmost point. In particular, one can check that, in any such measure,
the rightmost point is Gumble distributed.

The simplest point measure with the property~\eqref{supprop} is
the Poisson process with an exponential density $K e^{-x}$, for an
arbitrary $K$. Furthermore, all the decorated measures of
section~\ref{sec:Deco} when~$\nu_1$ is a Poisson process with an
exponential density $K e^{-x}$ are also superposable measures.
A natural question is then: can any superposable point
measure be constructed as a decorated exponential Poisson process for
a well chosen decoration measure ? A more specific question would be: is
the limiting point measure of the branching Brownian motion in the
$m_t+\ln(AZ)$ frame can be constructed as a decorated exponential Poisson
process?

We conjecture that the answer to the last question is yes. Indeed, it
is known \cite{DerridaSpohn.88,ArguinBovierKistler.10} that if one
considers two points chosen at random among the rightmost points in
a branching Brownian motion at a large time~$t$, then the time one needs to
go back to find the most recent common ancestor of those two points is
either very short (of order 1) or very long (of the order of the age $t$ of
the system). This means that one can group the rightmost points into
families where two points belong to the same family if the branching event
that generated them occurred recently. The branching Brownian motion
could then be a decorated exponential Poisson process, where the
rightmost points of all the families are Poisson distributed with an
exponential density \cite{ArguinBovierKistler.10} and where the members
of a given family form the decoration. This interpretation helps to
understand a question raised by our previous work
\cite{BrunetDerrida.09}: the distances between points in the
branching Brownian motion are given by (\ref{asympdi},\ref{valuesdn}), but
if one considers only the rightmost point of each family, then the
distances are given by \eqref{diexpo} with $\alpha=1$, as in the GREM or in
the Ruelle cascade, see section~\ref{RuelleCascade}.

\section{Conclusion}

In the present work, we have shown that all the statistical properties of
the rightmost points in a branching Brownian motion can be obtained by
solving a front equation with a well-chosen initial condition. The
distribution of the positions of the rightmost points seen in the frame
$m_t$ (the average position of the rightmost) has a long time limit. The
properties of the limiting distribution can be expressed as the long time
delays of the traveling wave solution of the Fisher-KPP equation when one
varies the initial condition. This limiting distribution is however
modified if one considers the union of several branching Brownian motions.

If one considers, however, only the distances between the points, for
example if one looks at the distribution of all the positions of the
rightmost points seen in the frame of the rightmost one, one obtains at
large times another limiting distribution which does not depend on the
initial positions of the branching Brownian motions (as long as there are
finitely many of them). We called this property superposability.

In section~\ref{quantitative} we have measured a few properties of this
limiting distribution, and in section~\ref{sec:disc} we explained how our
results can be extended to more general branching random walks.
Lastly, in section~\ref{statmeas}, we argued that in the
Lalley and Sellke frame, the branching Brownian motion satisfies a stronger
version of the superposability property, see~\eqref{supprop}.

In the future, it would be interesting to 
characterize more precisely the limiting measure of the branching Brownian
motion and of the branching random walks to see whether some universal
properties emerge. For example, we believe that the average density seen
from the rightmost point should always grow as in~\eqref{genaexpa}. It
would also be interesting to check whether, in Lalley's and Sellke's
frame, the limiting distribution is indeed a decorated exponential, and to
determine the properties of the decorating measure.

The question of ergodicity, raised at the beginning of
section~\ref{statmeas}, is also an interesting open question.

\begin{acknowledgements}
We would like to thank Julien Berestycki and Simon Harris for interesting
discussions.
\end{acknowledgements}

\appendix

\section{Numerical simulations}
\label{sec:numeric}

We performed the numerical simulations by discretizing the branching
Brownian motion in space
(with a grid length $\delta$) and time (with steps $s$) as in the
example~D) of section~\ref{sec:disc}.
The corresponding front equation is given by \eqref{ex5}, which is of
course the most straightforward discretization of the Fisher-KPP
equation~\eqref{FKPPH}. The solutions to the discrete
equation~\eqref{ex5} converge to the solutions of the Fisher-KPP
equation~\eqref{FKPPH}, if $s\to0$ with
$\delta/s^2$ held constant and small enough.

We used three sets of values for $s$ and $\delta$ and computed the
exact asymptotic velocity $v^*$ of the front and the decay rate $\gamma^*$ of
the asymptotic shape $F(x)$ by minimizing numerically the
function $v(\gamma)$ given in (\ref{vgammas}D). The decay
rate $\beta$ for the probability of observing a distance $a$
between the two rightmost points, see equation~\eqref{decayratedist2},
was computed using the recipe given at the end of
section~\ref{sec:distances} and explained in
appendix~\ref{sec:largedeviation}. All these values are
presented in table~\ref{tabval}
\begin{table}[ht]
\setlength{\extrarowheight}{2pt}
\centering
\begin{tabular}{|ll|lll|}
\hline
\multicolumn{1}{|c}{$s$} & 
\multicolumn{1}{c|}{$\delta$} & 
\multicolumn{1}{c}{$v^*$} &
\multicolumn{1}{c}{$\gamma^*$} & 
\multicolumn{1}{c|}{$\beta$}\\
\hline
0.25 & ${s^2}/5=0.0125$ 	& 1.980480133 & 1.004581693 & 2.387337826 \\
0.10 & ${s^2}/5=0.002 $	& 1.996840367 & 1.000747277 & 2.409772891\\
0.05 & ${s^2}/6\simeq0.000416667$& 1.999375296 & 1.000104046 & 2.412897517\\
\hline
\multicolumn{2}{|c|}{Fisher-KPP}		& 2	 & 1	 &
$\sqrt2+1=2.414213562$\\
\hline
\end{tabular}
\caption{Values of $v^*$, $\gamma^*$ and $\beta$ for our discretized
branching processes, compared to the values in the Fisher-KPP case.}
\label{tabval}
\end{table}

\begin{figure}[!ht]
\centering
\includegraphics[width=.45\textwidth]{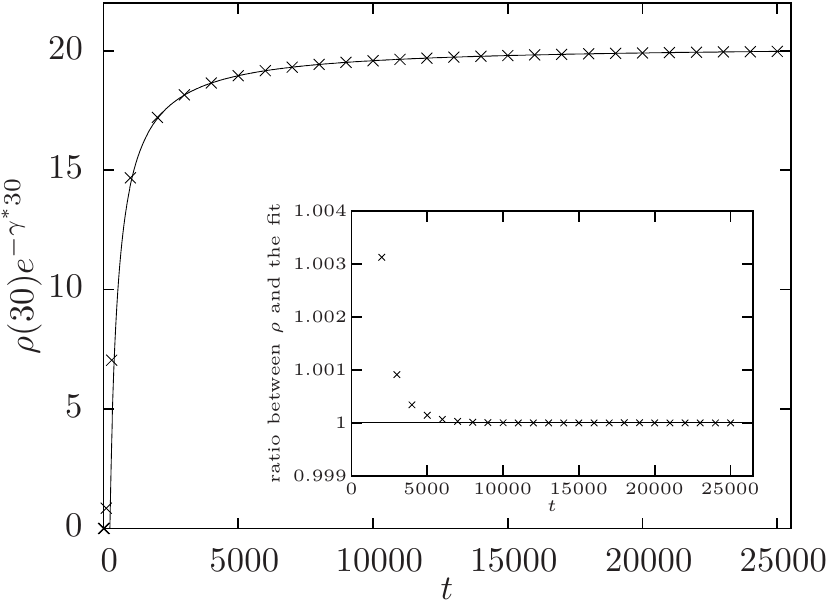}
\caption{Average density at a distance 30 from the rightmost point as a
function of time (symbols), fitted for $t\ge5000$ by the function
$A+B/t+C/t^{3/2}$ (line). The inset shows the quality of the fit by
displaying the ratio between the data points and the fitting function.}
\label{fit}
\end{figure}

The simulations were made on a finite but large domain centered
around the position of the front; typically it extended to about a
distance 1000 ahead and behind the center of the front (respectively 4000,
10000 or 20000 lattice sites depending on $s$). The values at $x=\pm\infty$
were exactly computed and used for the boundaries of the domain. Whenever
the front moved by more than one unit space lattice, the whole data set
was recentered. The simulations were performed up to large times of order
$10000$ (0.8 to 24 millions of time steps) and the data was extrapolated
to obtain a value at $t=\infty$. To do this extrapolation, we used a more
precise asymptotic expansion of the position of the front than
\eqref{Bramson}: according to \cite{EbertvanSaarloos.00},
\begin{equation}
m_t=v^* t-\frac3{2\gamma^*}\ln t +
\text{Constant}+\frac{a_{1/2}}{t^{1/2}}+\frac{a_1}t+\frac{a_{3/2}}{t^{3/2}}+\cdots,
\end{equation}
where the number $a_{1/2}$ does not depend on the initial condition. As
we measure the delay $m_t-m_t^{(\phi)}$, many terms cancel and one gets
\begin{equation}
m_t-m_t^{(\phi)}=f[\phi]+\frac{\delta a_1}t+\frac{\delta
a_{3/2}}{t^{3/2}}+\cdots.
\label{deltaasympt}
\end{equation}
All the quantities we measure are derivatives of $f[\phi]$, see
\eqref{the-frame-3}, and have therefore the same large time expansion as
\eqref{deltaasympt}. Thus, we extrapolated our numerical data to the large
time limit by fitting it with the function~$A+B/t+C/t^{3/2}$ for times larger
than (typically) 5000, see figure~\ref{fit}, and by using $A$ as the end
result.

On figure~\ref{fig:distrib_d}, the three data points were presented together;
on figure~\ref{fig:density}, we have drawn together for each data set the
function $\rho(X_1-a)e^{-\gamma^* a}$ using in each case the value of
$\gamma^*$ of table~\ref{tabval}. In both cases, the superposition was
nearly perfect, and so we expect that on the scales of the figure, the
curves would not change noticeably for smaller values of $s$ and $\delta$.

\section{Distribution of the distance between the two rightmost points}
\label{sec:largedeviation}

In this appendix we generalize, to any branching random walk, the argument
leading to the asymptotic decay \eqref{decayratedist2} of the distribution
of the distance between the two rightmost points in the branching Brownian
motion. 

We consider a generic branching random walk in discrete space
(with spacing $s$) and time (with intervals $\delta$)
defined by the following family of functions
\begin{equation}
p_n(r_1,\ldots,r_n)=\left(\,\parbox{7cm}{The probability that a point at
position $x$ branches during a time step into $n$ points located at
positions $x+r_1$, \ldots, $x+r_n$.}\,\right).
\label{pnr}
\end{equation}
We assume that $p_0=0$, so that there is no extinction. Then $p_1(r)$ can
be thought as the probability that the point does not branch but moves by
a distance $r$. The continuous time and/or space cases can be obtained as
suitable $s\to0$ and/or $\delta\to0$ limits.

Let $\exp[ t g(\beta) ]$ be the generating function of the displacement
during time $t$ of one point \emph{conditioned on the fact that this
point does not branch}:
\begin{equation}
e^{t g(\beta)} 
 = \sum_r e^{\beta r} \proba(\text{the point moves a distance $r$
 without branching
 during time $t$}).
\label{defgb}
\end{equation}
As the time steps are independent, the function $g(\beta)$ can be computed
during the time interval $\delta$ which gives
\begin{equation}
e^{\delta \,g(\beta)} = \sum_r p_1(r) e^{\beta r}.
\label{giveg}
\end{equation}
Note that $g(0)<0$ as soon as the branching probability is non-zero.
We want now to evaluate the probability that a point moves a distance $r$,
without branching, during time $t$. For large $t$, it takes the form 
\begin{equation}
\proba(\text{the point moves a distance $r$ without branching during time $t$})
\sim \exp\left[ t f\left(\frac r t\right)\right],
\end{equation}
where $f(c)$ is a large deviation function. Using \eqref{defgb}, one finds
that $f(c)$ and $g(\beta)$ are related by a Legendre transform
\begin{equation}
\begin{cases}
\beta=-f'(c),\\
g(\beta)=f(c)+\beta c.
\end{cases}
\label{largedev}
\end{equation}

Now, assuming as in section~\ref{sec:distances} that the events which
contribute most to a large distance $a$ between the two rightmost points
are those where the rightmost point moves, without branching, a distance
$a$ ahead of the frontier of the branching Brownian motion, one gets
\begin{equation}
P_{12}(a)\sim \max_{\tau} \left\{
\exp\left[\tau f\left(\frac{a+v^* \tau}{\tau}\right)\right] \right\},
\label{prob2}
\end{equation}
where $v^*$ is the velocity of the front.
For large $a$, the optimal $\tau$ is also large and it satisfies, by
derivation,
\begin{equation}
f\left(\frac a \tau + v^*\right) - \frac a \tau f'\left(\frac a
\tau + v^*\right)=0.
\label{fderiv}
\end{equation}
Let $c=a/\tau+v^*$.
Using \eqref{largedev}, equation \eqref{fderiv} becomes
\begin{equation}
g(\beta)=\beta v^*.
\label{whichbeta}
\end{equation}
Remarkably, this equation does not depend on $a$. Replacing into
\eqref{prob2} gives
\begin{equation}
P_{12}(a) \sim e^{-\beta a}.
\label{probdisc}
\end{equation}
The asymptotic decay rate of the probability distribution function of the
distance between the two rightmost points is therefore simply the positive
solution $\beta$ of~\eqref{whichbeta} with $g(\beta)$ given
by~\eqref{giveg}. In the branching Brownian motion, $g(\beta)=\beta^2-1$,
$v^*=2$, so that \eqref{whichbeta} gives indeed $\beta=1+\sqrt2$.

As can be checked easily from (\ref{giveg}), the function $g(\beta)$ is
convex. Therefore, as $g(0)<0$, equation \eqref{whichbeta} has at most one
positive solution and at most one negative solution. The positive solution
is the relevant one here.

We are now going to show that the negative solution of (\ref{whichbeta}),
if it exists, gives the asymptotic shape of the associated traveling wave.
We write the front equation associated to the branching point process
\eqref{pnr}. Using the same method as in section~\ref{sec:BBMFKPP}, we find
\begin{equation}
H_\phi(x,t+\delta)=\sum_{n\ge1} \sum_{r_1,\ldots,r_n}
p_n(r_1,\ldots,r_n)\prod_{i=1}^n H_\phi(x-r_i).
\label{eqpnr}
\end{equation}
We look at the shape $F$ of the traveling wave solution for a step
initial condition, which moves asymptotically at the velocity~$v^*$:
\begin{equation}
H_\phi(x,t)=F(x-v^* t).
\end{equation}
Using \eqref{eqpnr}, we see that in the \emph{stable} region $F(x)\ll1$
one has
\begin{equation}
F(x-v^*\delta)=\sum_r p_1(r) F(x-r) + {\cal O} (F^2).
\label{Fvstable}
\end{equation}
We look for an exponential solution to this linearized equation:
$F(x)\simeq
e^{\lambda x}$ with $\lambda>0$, as $F(-\infty)=0$. (Note that a periodic
modulation of this exponential could occur as $r$ takes only discrete
values.) Inserting into \eqref{Fvstable} and using \eqref{giveg}, one finds
that the equation for $\lambda$ is
\begin{equation}
g(-\lambda)=-\lambda v^*,
\end{equation}
which is the same equation as \eqref{whichbeta} for $\lambda=-\beta$.

To summarize, a \emph{positive} solution to \eqref{whichbeta} gives the
exponential decay rate of the probability distribution of the distance
between the two rightmost points, see \eqref{probdisc}, while
a \emph{negative} solution 
gives the coefficient
$-\lambda$ governing the shape of the front $F(x)$ in the stable
region $F(x)\ll1$.

\section{Lalley's and Sellke's result} \label{apxLS}

Lalley's and Sellke's theorem \cite{LalleySellke.87} is
\begin{equation}
\lim_{s\to\infty}
\lim_{t\to\infty}
\proba\big(X_1(t)<m_{t}+x \big| \{X_i(s)\}\big)
=
\exp\big(- A Z e^{-x}\big),
\label{LS}
\end{equation}
where $A$ is the constant appearing \cite{Bramson.83} in the large $x$
expansion of the function $F(x)$ defined in~\eqref{defF2}
\begin{equation}
F(x)\simeq 1- (A x + B) e^{-x}\qquad\text{for large $x$},
\label{asymptF}
\end{equation}
and $Z$ is defined in \eqref{defZ}. In words, given the positions
$\{X_i(s)\}$ at time $s$, there is a $t\to\infty$ limit to the probability
that the rightmost is on the left of $m_t+x$ which depends, obviously, on
the $\{X_i(s)\}$ and is as such a random variable. As $s$ goes to infinity,
this random variable converges almost surely to the Gumble distribution
around $\ln(AZ)$.

This result can be extended into the following: for any suitable function
$\phi$ (see section~\ref{BBM-frame}), one has
\begin{equation}
\lim_{s\to\infty}
\lim_{t\to\infty}
\Big\langle \prod_i \phi\big[m_t+x-X_i(t)\big] \Big| \{X_i(s)\}\Big\rangle
=\exp\big({-A Z e^{-x-f[\phi]}}\big),
\label{LS2}
\end{equation}
where $f[\phi]$ is the delay function \eqref{delay}. For $\phi=\theta$,
\eqref{LS2} reduces to \eqref{LS}.

We first give an outline of Lalley's and Sellke's proof applied to the
case~\eqref{LS2}. Given the positions $X_i(s)$ of the points at time~$s$,
the system as time~$t>s$ can be seen as a collection of independent
branching Brownian motions at time $t-s$ starting from the $X_i(s)$.
Therefore
\begin{equation}
\Big\langle \prod_i \phi\big[x-X_i(t)\big] \Big| \{X_i(s)\}\Big\rangle
	=\prod_i H_\phi\big(x-X_i(s),t-s\big),
\end{equation}
where the product in the right hand side is made on all the points present at
time~$s$.

We replace $x$ by $m_{t}+x$, to center around the position of the
front, and suppose $t$ large. It is easy to see from Bramson's
formula \eqref{Bramson} that $m_{t}=m_{t-s}+2s+{o}(1)$ as $t$ becomes
large, so that
\begin{equation}
\Big\langle \prod_i \phi\big[m_t+x-X_i(t)\big] \Big| \{X_i(s)\}\Big\rangle
	= \prod_i H_\phi\big(m_{t-s}+2s + x-X_i(s)+{o}(1),t-s\big),
\end{equation}
and, using \eqref{delay},
\begin{equation}
\lim_{t\to\infty} 
\Big\langle \prod_i \phi\big[m_t+x-X_i(t)\big] \Big| \{X_i(s)\}\Big\rangle
= \prod_i F\big(2s+x-X_i(s)+f[\phi]\big).
\end{equation}
We now take $s$ large. Of all the points present at time $s$, the rightmost
is around $2s-\frac32\log s$, see \eqref{Bramson}. Therefore, $2s-X_i(s)$
diverges for all $i$. Using \eqref{asymptF},
\begin{equation}
\lim_{t\to\infty} 
\Big\langle \prod_i \phi\big[m_t+x-X_i(t)\big] \Big| \{X_i(s)\}\Big\rangle
\simeq\exp\Big(-\sum_i
\big[A\big(2s+x-X_i(s)+f[\phi]\big)+B\big]e^{-2s-x+X_i(s)-f[\phi]}
\Big).
\end{equation}
Following Lalley and Sellke, we introduce the quantities
\begin{equation}
Y_s=\sum_i e^{-2s+X_i(s)},\qquad Z_s=\sum_i
\left[2s-X_i(s)\right]e^{-2s+X_i(s)},
\end{equation}
see \eqref{defZ}, so that
\begin{equation}
\lim_{t\to\infty} 
\Big\langle \prod_i \phi\big[m_t+x-X_i(t)\big] \Big| \{X_i(s)\}\Big\rangle
\simeq\exp\left(-
\left[A Z_s +\big(Ax+Af[\phi]+B\big)Y_s\right]e^{-x-f[\phi]}\right).
\label{lstLS}
\end{equation}
Finally, the most technical part of Lalley's and Sellke's proof is that $Y_s$ and $Z_s$ are
martingales converging when $s\to\infty$ to $\lim_{s\to\infty} Y_s=0$ and $\lim
_{s\to\infty}Z_s=Z>0$ respectively, which leads to \eqref{LS2}. We do not
reproduce this part of the proof here as it does not concern our extension
with the function $\phi$ and it works in \eqref{LS2} exactly as in
\eqref{LS}.

In (\ref{LS2}), the average is made on all the realizations
with a given set $\{X_i(s)\}$ of points at a large time $s$ but the only
relevant quantity appearing in the generating function \eqref{lstLS} is
$Z_s$. One would obviously have reached the same result if one had
conditioned by $Z_s$ instead of by the $\{X_i(s)\}$. Furthermore, 
as $Z_s$ converges quickly to $Z$, as illustrated on figure~\ref{fig2BBM},
we argue that conditioning by $Z_s$ at a large time $s$ or directly
conditioning  by $Z$ should be equivalent, hence
(\ref{LSint},\ref{LS2int}).


\end{document}